\documentclass[useAMS,usenatbib]{mn2e}

\usepackage[linkcolor=blue,colorlinks=true,breaklinks,citecolor=blue,urlcolor=blue]{hyperref}
\usepackage{amssymb,graphicx,threeparttable}

\title[UV+IR SFRs of Compact Group Galaxies]{The Ultraviolet and Infrared Star Formation Rates of Compact Group Galaxies: An Expanded Sample}

\author[Lenki\'{c} et al.]{Laura Lenki\'{c}$^{1}$\thanks{E-mail: llenkic2@uwo.ca}, Panayiotis Tzanavaris$^{2,3,4}$, Sarah Gallagher$^{1,5}$, Tyler Desjardins$^{1,6}$,
\newauthor
Lisa May Walker$^{7,8}$, Kelsey Johnson$^{7}$, Konstantin Fedotov$^{1}$, Jane Charlton$^{9}$, 
\newauthor
Ann Hornschemeier$^{2}$, Pat Durrell$^{10}$ and Caryl Gronwall$^{9}$\\
$^{1}$Department of Physics and Astronomy, University of Western Ontario, London, ON N6A 3K7, Canada\\
$^{2}$Laboratory for X-Ray Astrophysics, NASA Goddard Space Flight Center, Greenbelt, MD 20771, USA\\
$^{3}$Department of Physics, Joint Centre for Astrophysics, University of Maryland Baltimore County, 1000 Hilltop Cirlce, Baltimore, MD 21250, USA\\
$^{4}$Department of Physics and Astronomy, The Johns Hopkins University, Baltimore, MD 21218, USA\\
$^{5}$Centre for Planetary and Space Exploration, University of Western Ontario, London, ON N6A 3K7, Canada\\
$^{6}$Department of Physics and Astronomy, University of Kansas, Malott room 1082, 1251 Wescoe Hall Drive, Lawrence, KS 66045, USA\\
$^{7}$Astronomy Department, University of Virginia, Charlottesville, VA 22904, USA\\
$^{8}$Steward Observatory, University of Arizona, Tucson, AZ 85721, USA\\
$^{9}$Department of Astronomy and Astrophysics, Pennsylvania State University, University Park, PA 16802, USA\\
$^{10}$Department of Physics and Astronomy, Youngstown State University, Youngstown, OH 44555, USA}

\begin{document}

\maketitle

\label{firstpage}

\begin{abstract}
Compact groups of galaxies provide insight into the role of low-mass,
dense environments in galaxy evolution because the low velocity
dispersions and close proximity of galaxy members result in frequent
interactions that take place over extended timescales. We expand the
census of star formation in compact group galaxies by
\citet{tzanavaris10} and collaborators with Swift UVOT, Spitzer IRAC
and MIPS 24 \micron\ photometry of a sample of 183 galaxies in 46
compact groups. After correcting luminosities for the contribution
from old stellar populations, we estimate the dust-unobscured star
formation rate (SFR$_{\mathrm{UV}}$) using the UVOT uvw2
photometry. Similarly, we use the MIPS 24 \micron\ photometry to
estimate the component of the SFR that is obscured by dust
(SFR$_{\mathrm{IR}}$).  We find that galaxies which are MIR-active
(MIR-``red''), also have bluer UV colours, higher specific star
formation rates, and tend to lie in H~{\sc i}-rich groups, while
galaxies that are MIR-inactive (MIR-``blue'') have redder UV colours,
lower specific star formation rates, and tend to lie in H~{\sc i}-poor
groups. We find the SFRs to be continuously distributed with a peak at
about 1 M$_{\odot}$ yr$^{-1}$ , indicating this might be the most
common value in compact groups. In contrast, the specific star
formation rate distribution is bimodal, and there is a clear
distinction between star-forming and quiescent galaxies. Overall, our
results suggest that the specific star formation rate is the best
tracer of gas depletion and galaxy evolution in compact groups.
\end{abstract}


\section{Introduction}

Galaxy neighbourhoods cover a large range of richness as determined by
the number of members, ranging from poor groups to rich clusters. A
sub-class of galaxy groups is the class of compact groups (CGs). CGs
are small concentrations of galaxies that occupy a small angular
extent on the sky, where ``compact'' specifically indicates a handful
of galaxies within a few projected galaxy radii of each other.

CGs are characterized by low velocity dispersions (about 200 km
s$^{-1}$), short crossing times (about 0.016$H_{0}^{-1}$ or 200 Myr)
and high galaxy densities (comparable to the cores of dense clusters;
\citealt{hickson92}). However, they are found in low galaxy density
environments \citep[a result of the isolation
  criteria;][]{montenegro01}.  These combined properties should
enhance the occurrence of strong gravitational interactions including
mergers happening over long timescales ($\gtrsim1$ Gyr). CGs are
therefore excellent candidates for studying multiple concurrent
interactions and their effects on galaxy evolution. In addition, CGs
are the only nearby structures that roughly approximate the complex
interaction environments of the earlier universe when galaxies were
undergoing hierarchical assembly \citep{baron87}. Finally, galaxies in
group environments can transition from star-forming to quiescent
before falling into rich clusters (``pre-processing'') through
quenching mechanisms like early major mergers, starvation, and tidal
stripping (e.g. \citealt{cortese06,dressler13}). These mechanisms are
observed in CGs and play important roles in galaxy evolution in these
systems.

\citet{hickson94} conducted a study of the morphologies of galaxies in
92 HCGs and combined their data with previous results published on
their morphologies, kinematics, radio fluxes, infrared fluxes and
optical colour information. Their results showed that 43 per cent of
galaxies in CGs showed morphological and/or kinematic distortions that
were indicative of strong interactions (including possible mergers)
between galaxies.  Their results also showed that \mbox{$\ge$ 32 per
  cent} of groups have three or more galaxies that show signs of
interactions. The results of this study and further detailed
Fabry-Perot analyses (see \citet{torres-flores14} and references
therein) confirmed the importance of using the unique environment of
CGs to study complex galaxy interactions and their consequences on
galaxy evolution.

\citet{montenegro01} conducted an analysis of H~{\sc i} gas content in
72 HCGs, and proposed an evolutionary scheme for CGs comprising three
phases after loose groups have contracted into compact groups. The
first phase consists of largely undisturbed H~{\sc i} distributions
and kinematics with most of the gas present in the disks of the
galaxies and at most small amounts of gas found in developing tidal
tails. In the second phase, a significant amount of H~{\sc i} gas is
still in galaxy disks, but 30--60 per cent of it forms tidal
features. In the final phase, most of the H~{\sc i} gas (if
detectable) has been stripped from the galaxy disks into tidal
features. Throughout these three phases, the redistribution of gas
fuels star formation.

Further evidence that the distribution of gas plays an important role
in the evolution of galaxies was provided by \citet{johnson07}
(hereafter J07). They performed a {\em Spitzer Space Telescope} study
of the mid-infrared properties of 12 HCGs including CGs in all three
phases of the evolutionary scheme proposed by
\citet{montenegro01}. They found that galaxies residing in the most
H~{\sc i}-gas rich groups are more actively star-forming (based on
their \textit{Spitzer} IRAC mid-infrared colours and luminosities),
implying that galaxies in gas-rich environments experience star
formation until the neutral gas is consumed or is stripped by
interactions. In mid-infrared colour space, the groups studied by J07
showed a gap between gas-rich and gas-poor groups that is sparsely
populated by individual galaxies.  This dearth in colour space is not
seen in samples of field galaxies \citep{gallagher08,walker10}.  This
gap was interpreted to indicate a rapid stage of evolution in galaxy
properties that results from the complex dynamical interactions
characteristic of CG environments. J07 also proposed an evolutionary
scheme relating group gas richness to the individual galaxy member
morphologies: Type I groups are gas-rich with spirals or irregulars as
the main class of galaxies; Type III groups are gas-poor and tend to
be populated by elliptical and/or lenticular (E/S0 galaxies).
Finally, Type II groups are in an intermediate state in which CGs are
in the midst of forming their last generation of stars while consuming
their gas.

\citet[][hereafter T10]{tzanavaris10} included the ultraviolet (UV)
regime to determine whether or not, and to what extent, this would
provide support to the evolutionary scheme based on mid-infrared (MIR)
and radio (H~{\sc i}) work. They compiled UV data from the {\em Swift}
UV/ Optical Telescope (UVOT) as well as mid-IR data from the {\em
  Spitzer Space Telescope} MIPS 24 \micron\ camera on a sample of 11
HCGs (41 galaxies).  The photometry from the UVOT images was used to
determine the dust-unobscured component of the star formation rate
($\mathrm{SFR}_{\rm UV}$) that when combined with the dust-obscured
SFR ($\mathrm{SFR}_{\rm IR}$ from the 24 \micron\ luminosities)
provided an estimate for the total SFR ($\mathrm{SFR}_{\rm TOTAL}$)
for the 41 galaxies in the sample.  T10 found that the specific star
formation rates (SSFRs) defined as the $\mathrm{SFR}_{\rm
  TOTAL}$/$M_{*}$, where $M_{*}$ is the stellar mass, showed a clear
bimodal distribution in the sample of 11 HCGs, indicating that
galaxies are either actively star-forming or almost entirely quiescent
(i.e., not star forming). This provided a physical explanation for the
empirically observed gap in the MIR colours. The results also showed
-- as expected -- that E/S0 galaxies typically have low SSFRs, while
spiral galaxies have higher SSFRs (see also \citealt{bitsakis10}).

\citet{walker10, walker12} conducted further work on the mid-IR
properties of CGs, using an expanded sample of 49 CGs. They found that
for the larger sample of CGs, the J07 ``gap'' (with no galaxies in
that region of colour-space) was more accurately characterized as a
mid-IR ``canyon'', which still showed a significant dearth of
galaxies. Comparing to the Local Volume Legacy and the {\em Spitzer}
Infrared Nearby Galaxies Survey (LVL-SINGS) galaxies (local sample of
field galaxies;\citealt{dale09}), interacting galaxies \citep{smith07}
and galaxies in the Coma cluster \citep{jenkins07}, they found that
other than the galaxies in the Coma infall region, no other sample of
comparison galaxies exhibited a similar deficit in galaxy numbers in
mid-IR colour space. Similar to CGs, infall regions of clusters are
known to be sites of active galaxy evolution \citep{lubin02}. Other
authors using the \textit{Wide Field Infrared Survey Explorer} (WISE)
show bimodalities in the WISE colours for galaxies in environments
other than CGs \citep{ko13,yesuf14,alatalo14}. However, the WISE bands
(3.4, 4.6, 12, and 22 \micron\ ) are not the same as the IRAC bands
(3.6, 4.5, 5.8, and 8.0 \micron\ ), and therefore the IRAC colours
contain information that is different from those of the WISE
colours. For example, the 8 \micron\ IRAC channel can cleanly pick out
7.7 \micron\ PAH emission. The 12 \micron\ WISE bandpass that covers
the 7.7 \micron\ PAH feature is very broad (7-17 \micron\ ), and thus
the relative contributions of possible sources of emission (e.g.,
PAHs, silicate absorption or emission, and thermal dust continuum) are
often ambiguous. Furthermore, the 12 \micron\ WISE bandpass includes
the 11.3 \micron\ neutral PAH feature, which does not trace star
formation.

\citet{walker13} addressed the question of where the mid-IR canyon
galaxies fall in \textit{optical} colour-magnitude diagrams (CMDs),
i.e. whether they fall along the red sequence (``red and dead'' or
reddened galaxies), blue cloud (actively star-forming galaxies), or
the so-called optical green valley (transition region between the blue
cloud and red sequence where star formation has recently ceased).
They found that the MIR canyon galaxies fall on the optical red
sequence. However, optical colours of mature galaxies are not always
sensitive to low SSFRs, and therefore a UV-optical colour study is
required to determine if canyon galaxies truly reside in the
UV-defined ``green valley'' \citep{wyder07, martin07, thilker10}.

More recently, \citet{bitsakis14} estimated the dust masses,
luminosities and temperatures of 120 galaxies in 28 HCGs. They do this
by fitting their UV to sub-mm spectral energy distributions, using the
models of \citet{daCunha08}. In addition, they also calculate SFRs,
stellar masses, and SSFRs. They find that red late-type and dusty
lenticular galaxies appear to be a transition population between
star-forming and quiescent galaxies. Their results are generally
consistent with the picture that the interactions in CGs profoundly
shape the evolution of galaxy members.

In this paper, we expand on the work of T10 by using new
\textit{Swift} data (PI: Tzanavaris) to analyze the expanded sample of
49 CGs (with 193 galaxies) compiled by \citet{walker13}. The primary
aim is to investigate whether the result of T10 on the connection
between the SSFRs and mid-IR colours still holds for a significantly
larger sample of galaxies. Further, we address the question as to
where the mid-IR CG galaxies of all types fall within the UV-optical
CMDs, and if the UV colours independently offer any insight into star
formation in CG galaxies.

The paper is structured as follows: Section 2 describes the data
selection and analysis, Section 3 reviews and discusses the results
found in this work and Section 4 summarizes the paper and includes the
conclusion.  Throughout this paper we use a cosmology of $H_{0}$ = 70
km s$^{-1}$ Mpc$^{-1}$, $\Omega_{M}$ = 0.3, and $\Omega_{\Lambda}$ =
0.7 \citep{spergel07}.

\section{Observations and Data Analysis}
\subsection{HCG Sample Selection}

The sample of CGs studied here is the same sample studied by \citet{
  walker12} of 49 CGs: 33 Hickson Compact Groups and 16 Redshift
Survey Compact Groups (RSCGs). \citet{hickson82} constructed a
catalogue of 100 CGs using three photometric selection criteria: a
minimum of four galaxies (subsequent spectroscopy reduced this to 92
HCGs with at least 3 accordant member redshifts; \citealt{hickson92})
within 3 magnitudes of the brightest galaxy, a limiting surface
brightness to define compactness ($\bar{\mu} < 26$ mag arcsec$^{-2}$),
and an encircling ring devoid of bright galaxies to ensure isolation.

The RSCG catalogue of 89 CGs was constructed by \citet{
  barton96}. They used an automated search, applying a ``friends-of-
friends'' algorithm to find groups in the magnitude-limited Center for
Astrophysics 2 redshift survey (CfA2; \citealt{huchra86}) and Southern
Sky Redshift Survey (SSRS2; \citealt{daCosta91}) catalogues.  This
algorithm used projected distances between galaxies ($\Delta$D $\le$
50 kpc), and their line-of-sight velocity differences ($\Delta$v $\le$
1000 km s$^{-1}$) to determine whether two galaxies belong to a
group. This methodology was adopted in order to eliminate biases as a
function of distance which are introduced by visually selecting
groups, as \citet{hickson82} had done. Selecting CGs based on their
redshift reduces the possibility of chance projections, however the
isolation criterion was not rigorously satisfied and therefore many
RSCGs are actually components of larger structures. Because of this,
RSCGs 21, 67 and 68 are excluded from our analysis as they are part of
clusters, thus reducing our sample to 46 CGs \citep{walker12}. The
RSCGs in our sample were therefore selected to have properties similar
to HCGs. In addition, \citet{walker10} chose to include only CGs that
are at a sufficiently low redshift (z $<$0.035) that the polycyclic
aromatic hydrocarbon features do not shift out of their rest-frame
IRAC bands to maintain their rest-frame positions in mid-IR colour
space. The 46 remaining CGs in our sample satisfy this
requirement. The coordinates and group velocities of each CG are given
in Table \ref{sample}, as well as the breakdown of group properties
such as group H~{\sc i} content.

\begin{table*}
\caption{HCG and RSCG Sample}
\label{sample}
\begin{tabular}{lcccccccccc}

\hline
\hline
Group ID & RA$^{\mathrm{a,b}}$ & Dec$^{\mathrm{a,b}}$ & Velocity$^{\mathrm{c}}$ & Luminosity Distance$^{\mathrm{d}}$ & E(B-V)$^{\mathrm{e}}$ & UV & IR & Optical$^{\mathrm{f}}$ & Group & Group\\
  &  &  & (km s$^{-1}$) & (Mpc) &  &  &  &  & H~{\sc i} Type$^{\mathrm{g}}$ & H~{\sc i} Type$^{\mathrm{h}}$ \\ \hline
HCG 2 & 00h31m30.0s & +08$^\circ$25$^\prime$54$^{\prime\prime}$ & 3991 & 57.6 & 0.036 & Y & Y & Y  & \MakeUppercase{\romannumeral 1} & Rich \\
HCG 4 & 00h34m16.0s & $-$21$^\circ$26$^\prime$48$^{\prime\prime}$ & 7764 & 113.0 & 0.018 & Y & Y & Y & \MakeUppercase{\romannumeral 2} & Rich \\
HCG 7 & 00h39m24.0s & +00$^\circ$52$^\prime$42$^{\prime\prime}$ & 3885 & 56.0 & 0.018 & Y & Y & Y & \MakeUppercase{\romannumeral 2} & Int \\
HCG 15    & 02h07m39.0s & +02$^\circ$08$^\prime$18$^{\prime\prime}$ & 6568 & 95.4 & 0.026 & Y & Y & Y & \MakeUppercase{\romannumeral 3} & Int \\
HCG 16    & 02h09m31.3s & $-$10$^\circ$09$^\prime$31$^{\prime\prime}$ & 3706 & 53.4 & 0.022 & Y & Y & Y & \MakeUppercase{\romannumeral 2} & Int \\
HCG 19    & 02h42m45.0s & $-$12$^\circ$24$^\prime$42$^{\prime\prime}$ & 3989 & 57.6 & 0.03 & Y & Y & N & \MakeUppercase{\romannumeral 1} & Rich \\
HCG 22    & 03h03m33.0s & $-$15$^\circ$41$^\prime$00$^{\prime\prime}$ & 2522 & 36.3 & 0.048 & Y & Y & Y & \MakeUppercase{\romannumeral 2} & Int \\
HCG 25    & 03h20m43.0s & $-$01$^\circ$03$^\prime$06$^{\prime\prime}$ & 6185 & 89.8 & 0.063 & Y & Y & Y & \MakeUppercase{\romannumeral 1} & Rich \\
HCG 26    & 03h21m54.0s & $-$13$^\circ$38$^\prime$48$^{\prime\prime}$ & 9319 & 136.0 & 0.053 & Y & Y & N & \MakeUppercase{\romannumeral 2} & Rich \\
HCG 31    & 05h01m38.3s & $-$04$^\circ$15$^\prime$25$^{\prime\prime}$ & 4026 & 58.1 & 0.045 & Y & Y & Y & \MakeUppercase{\romannumeral 1} & Rich \\
HCG 33    & 05h10m48.0s & +18$^\circ$02$^\prime$06$^{\prime\prime}$ & 7779 & 113.0 & 0.305 & Y & Y & N & \MakeUppercase{\romannumeral 2} & Rich \\
HCG 37    & 09h13m35.0s & +30$^\circ$00$^\prime$54$^{\prime\prime}$ & 6940 & 101.0 & 0.027 & Y & Y & Y & \MakeUppercase{\romannumeral 3} & Int \\
HCG 38    & 09h27m39.0s & +12$^\circ$16$^\prime$48$^{\prime\prime}$ & 9067 & 133.0 & 0.035 & Y & Y & Y & \MakeUppercase{\romannumeral 2} & Int \\
HCG 40    & 09h38m54.0s & $-$04$^\circ$51$^\prime$06$^{\prime\prime}$ & 7026 & 102.0 & 0.057 & Y & Y & N & \MakeUppercase{\romannumeral 2} & Int \\
HCG 42    & 10h00m22.0s & $-$19$^\circ$39$^\prime$00$^{\prime\prime}$ & 4332 & 62.6 & 0.038 & Y & Y & N & \MakeUppercase{\romannumeral 2} & Poor \\
HCG 47    & 10h25m48.0s & +13$^\circ$43$^\prime$54$^{\prime\prime}$ & 9843 & 147.0 & 0.038 & Y & Y & Y & \MakeUppercase{\romannumeral 2} & Int \\
HCG 48    & 10h37m45.0s & $-$27$^\circ$04$^\prime$48$^{\prime\prime}$ & 3163 & 45.6 & 0.063 & Y & Y & N & \MakeUppercase{\romannumeral 3} & $\cdots$ \\
HCG 54    & 11h29m15.5s & +20$^\circ$35$^\prime$06$^{\prime\prime}$ & 1797 & 25.8 & 0.018 & Y & Y & Y & \MakeUppercase{\romannumeral 2} & Rich \\
HCG 56    & 11h32m39.6s & +52$^\circ$56$^\prime$25$^{\prime\prime}$ & 8279 & 121.0 & 0.013 & Y & Y & Y & \MakeUppercase{\romannumeral 2} & Poor \\
HCG 57    & 11h37m50.0s & +21$^\circ$59$^\prime$06$^{\prime\prime}$ & 9436 & 138.0 & 0.028 & N & Y & N & \MakeUppercase{\romannumeral 3} & Poor \\
HCG 59    & 11h48m26.6s & +12$^\circ$42$^\prime$40$^{\prime\prime}$ & 4392 & 63.5 & 0.032 & Y & Y & Y & \MakeUppercase{\romannumeral 3} & Rich \\
HCG 61    & 12h12m25.0s & +29$^\circ$11$^\prime$24$^{\prime\prime}$ & 4186 & 60.4 & 0.019 & Y & Y & Y & \MakeUppercase{\romannumeral 2} & Rich \\
HCG 62    & 12h53m08.0s & $-$09$^\circ$13$^\prime$24$^{\prime\prime}$ & 4443 & 68.7 & 0.045 & Y & Y & N & \MakeUppercase{\romannumeral 3} & Poor \\
HCG 67    & 13h49m03.0s & $-$07$^\circ$12$^\prime$18$^{\prime\prime}$ & 7633 & 109.0 & 0.031 & N & Y & N & \MakeUppercase{\romannumeral 3} & Poor \\
HCG 68    & 13h53m40.9s & +40$^\circ$19$^\prime$07$^{\prime\prime}$ & 2583 & 37.1 & 0.01 & N & Y & N & \MakeUppercase{\romannumeral 2} & Int \\
HCG 71    & 14h11m04.0s & +25$^\circ$29$^\prime$06$^{\prime\prime}$ & 9240 & 135.0 & 0.018 & Y & Y & Y & $\cdots$ & $\cdots$ \\
HCG 79    & 15h59m11.9s & +20$^\circ$45$^\prime$31$^{\prime\prime}$ & 4439 & 64.1 & 0.048 & Y & Y & Y & \MakeUppercase{\romannumeral 2} & Rich \\
HCG 90    & 22h02m05.0s & $-$31$^\circ$58$^\prime$00$^{\prime\prime}$ & 2364 & 34.0 & 0.023 & Y & Y & N & \MakeUppercase{\romannumeral 3} & Poor \\
HCG 91    & 22h09m10.4s & $-$27$^\circ$47$^\prime$45$^{\prime\prime}$ & 6843 & 99.5 & 0.017 & N & Y & N & \MakeUppercase{\romannumeral 2} & $\cdots$ \\
HCG 92    & 22h35m59.0s & +33$^\circ$57$^\prime$30$^{\prime\prime}$ & 6119 & 88.8 & 0.07 & Y & Y & N & \MakeUppercase{\romannumeral 2} & Int \\
HCG 96    & 23h27m58.0s & +08$^\circ$46$^\prime$24$^{\prime\prime}$ & 8384 & 122.0 & 0.052 & Y & Y & N & \MakeUppercase{\romannumeral 2} & Rich \\
HCG 97    & 23h47m22.9s & $-$02$^\circ$19$^\prime$34$^{\prime\prime}$ & 6174 & 89.6 & 0.031 & Y & Y & Y & \MakeUppercase{\romannumeral 3} & Int \\
HCG 100    & 00h01m20.0s & +13$^\circ$08$^\prime$00$^{\prime\prime}$ & 4976 & 72.0 & 0.072 & Y & Y & Y & \MakeUppercase{\romannumeral 2} & Rich \\
RSCG 4    & 00h42m49.5s & $-$23$^\circ$33$^\prime$11$^{\prime\prime}$ & 6302 & 91.5 & 0.017 & Y & Y & Y & \MakeUppercase{\romannumeral 1} & Int \\
RSCG 6    & 01h16m12.0s & +46$^\circ$44$^\prime$18$^{\prime\prime}$ & 4847 & 70.1 & 0.072 & Y & N & Y & \MakeUppercase{\romannumeral 1} & Rich \\
RSCG 15    & 01h52m41.4s & +36$^\circ$08$^\prime$46$^{\prime\prime}$ & 4543 & 65.7 & 0.078 & Y & Y & N & \MakeUppercase{\romannumeral 3} & $\cdots$ \\
RSCG 17    & 01h56m21.8s & +05$^\circ$38$^\prime$37$^{\prime\prime}$ & 5415 & 78.4 & 0.046 & Y & Y & Y & \MakeUppercase{\romannumeral 2} & $\cdots$ \\
RSCG 21$^{\mathrm{i}}$    & 03h19m36.6s & +41$^\circ$33$^\prime$39$^{\prime\prime}$ & 4937 & 71.4 & 0.146 & Y & Y & Y & \MakeUppercase{\romannumeral 2} & $\cdots$ \\
RSCG 31    & 09h17m26.0s & +41$^\circ$57$^\prime$18$^{\prime\prime}$ & 2009 & 28.8 & 0.017 & Y & N & Y & \MakeUppercase{\romannumeral 1} & Int \\
RSCG 32    & 09h19m51.0s & +33$^\circ$46$^\prime$18$^{\prime\prime}$ & 6987 & 102.0 & 0.015 & Y & Y & Y & \MakeUppercase{\romannumeral 3} & Int \\
RSCG 34    & 09h43m12.0s & +31$^\circ$54$^\prime$42$^{\prime\prime}$ & 1765 & 25.3 & 0.018 & Y & Y & Y & \MakeUppercase{\romannumeral 3} & Int \\
RSCG 38    & 10h51m46.7s & +32$^\circ$51$^\prime$31$^{\prime\prime}$ & 1783 & 25.6 & 0.02 & Y & Y & Y & $\cdots$ & Rich \\
RSCG 42    & 11h36m51.3s & +19$^\circ$59$^\prime$19$^{\prime\prime}$ & 6624 & 96.2 & 0.023 & Y & N & Y & \MakeUppercase{\romannumeral 2} & Rich \\
RSCG 44    & 11h44m00.6s & +19$^\circ$56$^\prime$44$^{\prime\prime}$ & 6623 & 96.2 & 0.019 & Y & Y & Y & \MakeUppercase{\romannumeral 3} & Poor \\
RSCG 64    & 12h41m32.0s & +26$^\circ$04$^\prime$06$^{\prime\prime}$ & 5083 & 73.6 & 0.014 & Y & Y & Y & \MakeUppercase{\romannumeral 1} & Rich \\
RSCG 66    & 12h43m17.9s & +13$^\circ$11$^\prime$47$^{\prime\prime}$ & 1220 & 17.5 & 0.023 & Y & Y & N & $\cdots$ & $\cdots$ \\
RSCG 67$^{\mathrm{i}}$    & 12h59m32.8s & +27$^\circ$57$^\prime$27$^{\prime\prime}$ & 7464 & 109.0 & 0.008 & N & Y & N & $\cdots$ & $\cdots$ \\
RSCG 68$^{\mathrm{i}}$    & 13h00m10.7s & +27$^\circ$58$^\prime$17$^{\prime\prime}$ & 6864 & 99.8 & 0.008 & N & Y & N & $\cdots$ & $\cdots$ \\
RSCG 86    & 23h38m34.4s & +27$^\circ$01$^\prime$24$^{\prime\prime}$ & 8348 & 122.0 & 0.06 & Y & Y & Y & \MakeUppercase{\romannumeral 2} & $\cdots$ \\
\hline

\multicolumn{11}{p{\textwidth}}{$^{\mathrm{a}}$Position data taken from \citet{hickson82} for HCG sample. 
$^{\mathrm{b}}$Position data taken from \citet{barton96} for RSCG sample. 
$^{\mathrm{c}}$Velocities corrected to CMB taken from \url{http://ned.ipac.caltech.edu/} for all CGs in our sample. 
$^{\mathrm{d}}$CMB corrected values from \url{http://ned.ipac.caltech.edu/} with a cosmology of H$_{0}$ = 70 $\mathrm{km s^{-1} Mpc^{-1}}$, $\Omega_{\mathrm{M}}$ = 0.3, $\Omega_{\Lambda}$ = 0.7. 
$^{\mathrm{e}}$Values taken from \url{http://ned.ipac.caltech.edu/} with a cosmology of H$_{0}$ = 70 $\mathrm{km s^{-1} Mpc^{-1}}$, $\Omega_{\mathrm{M}}$ = 0.3, $\Omega_{\Lambda}$ = 0.7. 
$^{\mathrm{f}}$Sloan Digital Sky Survey Optical r band Data from \citet{walker13}. 
$^{\mathrm{g}}$Group types as defined by J07. 
$^{\mathrm{h}}$Walker et al. (2015), submitted. 
$^{\mathrm{i}}$These CGs are part of clusters. Their photometry is presented in Table \ref{phot}, but they are not included in the subsequent analysis.
}
\end{tabular}

\end{table*}

\subsection{UV Data}

{\em Swift} is a multi-wavelength observatory \citep{gehrels04} whose
primary goal is to study gamma-ray bursts. When {\em Swift} is not
following gamma ray burst targets, there is a fill-in program of other
observations. The ultraviolet optical telescope (UVOT)
\citep{roming05} instrument is a 30~cm telescope with a 17 $\times$ 17
square arcminute field of view that provides ultraviolet and optical
coverage (1600--8000 \AA) with a spatial resolution of $\sim$
2.5\arcsec. It uses a micro-channel-plate, intensified photon-
counting CCD detector. It has 256 $\times$ 256 active pixels, each of
which is subdivided into 8 $\times$ 8 pixels using a centroiding
algorithm \citep{breeveld10}.

\begin{table*}
\caption{UVOT Filter Characteristics}
\label{UVOT}
\begin{tabular}{lcccccc}

\hline
\hline
Filter & Conversion Factor$^{\mathrm{a}}$ & Error & Effective Wavelength & Width$^{\mathrm{b}}$ & Zero Point$^{\mathrm{c}}$ & Error \\
       & \multicolumn{2}{c}{Count Rates to Flux (erg s$^{-1}$ cm$^{-2}$)} & \AA & \AA &  &  \\ \hline
$uvw2$ & $6.2 \times 10^{-16}$  & $1.4 \times 10^{-17}$ & 2030 & 657 & 19.11 & 0.03 \\
$uvm2$ & $8.50 \times 10^{-16}$ & $5.6 \times 10^{-18}$ & 2231 & 498 & 18.54 & 0.03 \\
$uvw1$ & $4.00 \times 10^{-16}$ & $9.7 \times 10^{-18}$ & 2634 & 693 & 18.95 & 0.03 \\
$u$    & $1.63 \times 10^{-16}$ & $2.5 \times 10^{-17}$ & 3501 & 785 & 19.36 & 0.02 \\
\hline

\multicolumn{7}{p{\textwidth}}{$^{\mathrm{a}}$Data from \citet{poole08}. 
$^{\mathrm{b}}$For UVOT, these are the full width at half maximum values. 
$^{\mathrm{c}}$Data from \citet{breeveld11}, AB magnitudes.
}
\end{tabular}

\end{table*}

The data used in this study originated from ``fill-in'' observations
with UVOT's three UV filters ($uvw2$, $uvm2$, $uvw1$) as well as the
bluest optical filter ($u$). The characteristics of these filters are
given in Table \ref{UVOT}. All UV data (PI: Tzanavaris) were
downloaded from the {\em Swift}
archive\footnote{\url{http://heasarc.nasa.gov/docs/swift/archive/}}
and Table \ref{obs} gives a list of the observation IDs used for this
study.

\begin{table*}
\caption{Observation Log for Swift UVOT HCG and RSCG Data - The full table is available online}
\label{obs}
\begin{tabular}{lcccccc}

\hline
\hline
Group ID & Observation ID & Dates & \multicolumn{4}{c}{Total Exposure Time} \\
      &   &   & $uvw2$ & $uvm2$ & $uvw1$ & $u$ \\
      &   &   & (s) & (s) & (s) & (s) \\ \hline
HCG 2	& 00035906001 & 2007 Feb 11, 2007 Feb 12     & 2449   & 2431   & 1632   & 811     \\
	& 00035906002 & 2011 Nov 1		     &	      &        &        &         \\
HCG 4	& 00091109001 & 2011 May 5 		     & $\cdots$ & 1594   & 2916   & 5122    \\
	& 00091109002 & 2011 May 22 		     &        &        &        &         \\
	& 00091109003 & 2001 May 26, 2011 May 27     &        &        &        &         \\
HCG 7   & 00035907001 & 2006 Oct 28, 2006 Oct 30     & 4236   & 4633   & 2867   & 1406    \\
	& 00035907002 & 2006 Nov 10		     &        &        &        &         \\
	& 00035907003 & 2007 Jan 29                  &        &        &        &         \\
	& 00035907004 & 2007 Feb 10, 2007 Feb 11     &        &        &        &         \\ 
\hline

\end{tabular}
\end{table*}

The data were reduced following standard procedures and dedicated UVOT
analysis routines which are part of the HEASOFT package (version
6.15.1) (see T10 for more details). In brief, sky images were prepared
from raw images and event files, and were aspect-corrected. Final
exposure maps and images were made by combining separate exposures for
a given observation. The pixel scale of these images is 0.502 arcsec
pixel$^{-1}$.

The photometric zero points for converting UVOT count rates to the AB
magnitude system \citep{oke74} are provided by the Goddard Space
Flight
Center\footnote{\url{http://swift.gsfc.nasa.gov/analysis/uvot\_digest/zeropts.html}}
(see also \citealt{breeveld11}). The zero points are: $19.11 \pm
0.03$, $18.54 \pm 0.03$, $18.95 \pm 0.03$ and $19.36 \pm 0.02$
$\mathrm{mag}$ for the $uvw2$, $uvm2$, $uvw1$ and $u$ filters
respectively (see Table \ref{phot}).

\subsubsection{$uvm2$ AB Magnitudes}

To compare UV-optical colours of CGs to other galaxy samples in the
literature that use GALEX $NUV-r$ (e.g., \citealt{hammer12}), we
convert $uvm2$ fluxes to AB magnitues. In terms of effective
wavelength, the $uvm2$ filter most closely matches the GALEX NUV.

\begin{equation}
m_{uvm2} \equiv -2.5 \times \log_{10}(C_{uvm2}) + {\rm ZP}_{uvm2}
\label{mag}
\end{equation}

where $C_{uvm2}$ is the $uvm2$ filter count rate and ZP$_{uvm2}$ is
the zero point of the $uvm2$ filter (Table \ref{UVOT}).

\subsubsection{Coincidence Loss}

UVOT is a photon-counting detector and thus suffers from coincidence
loss at high count rates \citep{fordham00} because more than one
photon may be registered within a single readout
interval. \citet{poole08} studied coincidence loss in 5{\arcsec}
radius apertures by comparing observed count rates to theoretical
ones. They determined that for such apertures, coincidence loss
becomes an important effect at $\approx$10 counts s$^{-1}$. They
estimate that the true flux for such a count rate is 5 per cent higher
(see T10 for more details). \citet{lanz13} further discuss that
coincidence loss becomes greater than 1 per cent when the count rate
is greater than 0.007 counts per second per pixel. They correct for
coincidence loss in extended sources by excluding regions of high
count rates, and then measuring them as point sources and applying the
corrections of \citet{poole08}.

We thus checked our UV data for coincidence loss effects using the
method of \citet{poole08}, by measuring the non-background subtracted
count rates in circular apertures within a 5{\arcsec} radius centered
on the surface brightness peak. Consistent with the results of T10, we
found that the {\em u} filter images were most significantly affected
by coincidence loss; 32 of 120 galaxies with {\em u} data suffered
from coincidence loss, whereas only 6, 2, and 5 galaxies suffered from
coincidence loss effects in the {\em uvw1}, {\em uvm2}, and $uvw2$
filters, respectively. Data thus affected are indicated in Table
\ref{phot} with a superscript and table footnote.

\subsection{Mid-Infrared Data}

The \textit{Spitzer} Infrared Array Camera (IRAC; \citealt{fazio04})
is a four-channel camera with a $5'.2 \times 5'.2$ field of view that
has imaging capabilities at 3.6, 4.5, 5.8, and 8.0 \micron. The IRAC
detector arrays have a size of $256 \times 256$ pixels and the images
have a pixel scale of 1.2 arcsec pixel$^{-1}$. The images for our
sample of CGs are archival data presented by \citet{walker12}.

The Multiband Imaging Photometer for {\em Spitzer} (MIPS;
\citealt{rieke04}) is an instrument that provides three imaging bands
at 24, 70, and 160 \micron. At 24 \micron, the instrument has a
resolution of $\sim$ 6{\arcsec} \citep{dole06}. {\em Spitzer} MIPS (24
\micron) data were obtained from the Spitzer Heritage Archive
\footnote{\url{http://archive.spitzer.caltech.edu/}}. The basic
calibrated data were downloaded and processed through the template
overlap and mosaic pipelines of the \textit{Spitzer} MOPEX package
(version 18.5.6, \citealt{makovoz06}). These images have a pixel scale 
of 2.45 arcsec pixel$^{-1}$.

\subsection{Optical Data} 

We use Sloan Digital Sky Survey (SDSS) $r$-band magnitudes to create
UV-optical colour-magnitude diagrams. The optical photometry was
obtained from \citet{walker13} who performed custom photometry on SDSS
images using SURPHOT \citep{reines08}.  This IDL routine determines
apertures based on the contour levels of a reference image, and then
applies those apertures to every image. \citet{walker13} used the sum
of the $gri$ images of each galaxy as the reference image. Not all CGs
in our sample have optical photometry; the data status is indicated in
column 9 of Table~\ref{sample}.

\subsection{Extended Source Photometry}

Before photometry was performed, all UVOT and IRAC images were
convolved with a Gaussian kernel to the MIPS 24 \micron\ point spread
function (PSF), which is significantly broader than the UVOT PSF
($\sim$ 2.5{\arcsec} vs.  $\sim$ 6{\arcsec} for MIPS).

The 3.6 \micron\ fluxes primarily probe the distribution of stellar
mass in a galaxy, and they are intermediate in wavelength between the
UV and 24 \micron\ data.  We thus used the convolved images in this
band to define common photometric Kron apertures for each galaxy with
SExtractor (version 2.8.6, \citealt{bertin96}). We chose the default
values for all SExtractor parameters, except the ones listed in Table
\ref{params}. The latter were empirically established as the best
values for obtaining well-defined apertures for individual galaxies
given the small projected separations of CG galaxies.

\begin{table}
\caption{SExtractor Parameters}
\label{params}
\begin{tabular}{lc}

\hline
\hline
Parameter & Value \\
\hline
DETECT\_MINAREA & 10 \\
DETECT\_THRESH & 10 \\
ANALYSIS\_THRESH & 10 \\
DEBLEND\_NTHRESH & 20 \\
DEBLEND\_MINCONT & 0.001 \\
SEEING\_FWHM & 1.66 \\
\hline

\end{tabular}
\end{table}

For each galaxy, background annuli were defined between 1.5 and 2
times the Kron radius of the galaxy aperture, where the Kron radius is
a luminosity-weighted radius which will include more than $\approx$ 90
per cent of a galaxy's light \citep{kron80}. All the apertures were
visually inspected and adjusted as needed to avoid overlaps between
adjacent galaxies. In some cases, the angular separation of the
galaxies was too small (or non-existent) to separate them, and we
combined the apertures for these. There are six CGs where this was the
case and these appear in Table \ref{phot} with a single entry for the
combined galaxies.  The name includes the designations of all galaxies
in the aperture (e.g., HCG 31ACE has a single aperture encompassing
the three galaxies HCG 31A, 31C, and 31E).

Photometry was then carried out on all UVOT, IRAC 3.6 and 4.5
\micron\ and MIPS 24 \micron\ images, with our apertures, using the
IDL photometry tool {\tt SURPHOT} \citep{reines08} modified to
accommodate photon-counting data (T10). The results of the UV and IR
photometry are presented in Table \ref{phot}. Every galaxy in our
sample with \textit{Swift} data was detected in the UV, however there
were 2 non-detections in the 24 \micron\ images (HCG 59B and HCG 97C).

\begin{table*}
\caption{UVOT, MIPS 24 $\mu$m and IRAC 3.6 and 4.5 $\mu$m Fluxes for HCG and RSCG Galaxies. The full table is available online.}
\label{phot}
\begin{tabular}{lrrrrrrcrrr}

\hline
\hline
CG ID & \multicolumn{5}{c}{UVOT Flux Densities} &   & \multicolumn{4}{c}{\textit{Spitzer} Flux Densities} \\
   & $uvw2$ & Corrected$^{\mathrm{a}}$ & $uvm2$ & $uvw1$ & $u$ & m$_{uvm2}$$^{\mathrm{b}}$ & 3.6 $\mu$m & 4.5 $\mu$m & 24 $\mu$m & Corrected$^{\mathrm{a}}$ \\
   & \multicolumn{5}{c}{($10^{-16}$ erg s$^{-1}$ cm$^{-2}$)} &   & \multicolumn{4}{c}{(mJy)} \\ \hline 
HCG 2A	& $328.3 \pm 12.9$ & $204.0$ & $283.8 \pm 9.6$ & $188.8 \pm 7.6$ & $159.1 \pm 4.3$$^{\mathrm{c}}$ & 15.1 & $18.4 \pm 1.8$ & $12.5 \pm 1.2$ & $127.3 \pm 12.7$ & $125.4 \pm 12.4$ \\
HCG 2B	& $124.6 \pm 5.2$ & $75.8$ & $115.1  \pm 4.4$ & $73.2  \pm 3.1$ & $70.5  \pm 2.0$$^{\mathrm{c}}$ & 16.0 & $15.5 \pm 1.5$ & $11.2 \pm 1.1$ & $344.2 \pm 34.4$ & $342.6 \pm 34.2$ \\
HCG 2C	& $85.8 \pm 3.7$ & $52.2$ & $78.6 \pm 3.2 $ & $49.5 \pm 2.2$ & $54.9 \pm 1.7$ & 16.5 & $10.4 \pm 1.0$ & $6.7 \pm 1.3$ & $22.8 \pm 2.4$ & $21.8 \pm 2.2$ \\
HCG 4A	& $\cdots$ & $\cdots$ & $168.8 \pm 6.0$ & $136.0 \pm 5.4$ & $132.8 \pm 3.4$$^{\mathrm{c}}$ & 15.5 & $37.4 \pm 3.7$ & $27.5 \pm 2.8$ & $402.1 \pm 40.2$ & $398.4 \pm 39.8$ \\
HCG 4B  & $\cdots$ & $\cdots$ & $23.3 \pm 1.2$ & $19.2 \pm 0.9$ & $21.4 \pm 0.6$ & 17.6 & $4.4 \pm 2.2$ & $2.7 \pm 1.3$ & $12.5 \pm 1.3$ & $12.1 \pm 1.2$ \\
HCG 4C  & $\cdots$ & $\cdots$  & $2.8 \pm 0.4$  & $4.7 \pm 0.3$ & $10.1 \pm 0.3$ & 19.9 & $7.5 \pm 1.5$ & $5.2 \pm 1.0$ & $2.8 \pm 1.4$ & $2.1 \pm 1.1$ \\
HCG 4D  & $\cdots$ & $\cdots$ & $15.0 \pm 0.9$ & $13.0 \pm 0.7$ & $13.4 \pm 0.4$ & 18.1 & $4.3 \pm 2.1$ & $2.8 \pm 1.4$ & $29.4 \pm 2.9$ & $29.0 \pm 2.9$ \\
HCG 7A  & $59.7 \pm 2.5$ & $29.6$ & $51.6 \pm 2.0$ & $67.2 \pm 2.8$ & $115.8 \pm 3.1$$^{\mathrm{c}}$ & 16.7 & $72.9 \pm 7.3$ & $47.1 \pm 4.7$ & $294.5 \pm 29.4$ & $287.2 \pm 28.7$ \\
HCG 7B  & $15.5 \pm 0.8$ & $4.7$ & $12.0 \pm 0.7$ & $26.5 \pm 1.2$ & $63.3 \pm 1.8$$^{\mathrm{c}}$ & 18.3 & $34.7 \pm 3.5$ & $21.7 \pm 2.2$ & $11.2 \pm 1.1$ & $7.7 \pm 1.5$ \\
HCG 7C  & $165.4 \pm 6.4$ & $113.0$ & $145.7 \pm 4.9$ & $123.7 \pm 5.0$ & $141.0 \pm 3.8$ & 15.6 & $36.4 \pm 3.6$ & $23.7 \pm 2.4$ & $78.9 \pm 7.9$ & $75.3 \pm 7.5$ \\
HCG 7D  & $43.1 \pm 1.9$ & $29.2$ & $39.9 \pm 1.6$ & $33.6 \pm 1.5$ & $44.5 \pm 1.3$ & 17.0 & $10.8 \pm 1.1$ & $7.0 \pm 1.4$ & $11.3 \pm 1.1$ & $10.2 \pm 1.0$ \\
\hline

\multicolumn{11}{p{\textwidth}}{$^{\mathrm{a}}$$uvw2$ and 24 $\mu$m values corrected for emission from old stellar populations (see section 2.5). 
$^{\mathrm{b}}$AB magnitudes for the $uvm2$ filter. 
$^{\mathrm{c}}$These are the galaxies which experienced coincidence loss in the $u$ filter.
}
\end{tabular}

\end{table*}

\begin{figure}
\begin{center}
\includegraphics[width = \columnwidth]{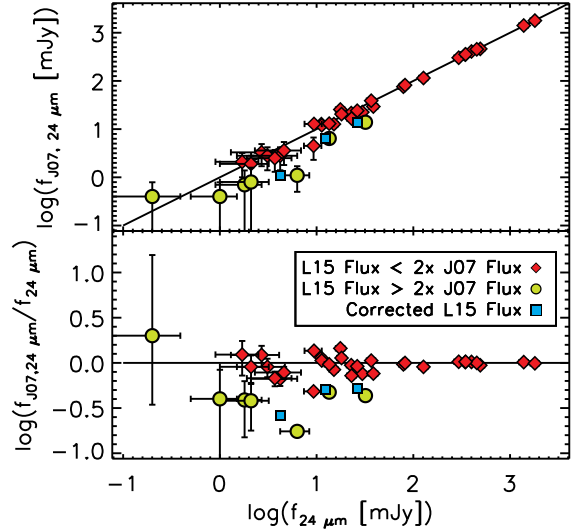}
\caption{Top: J07 and T10 24 \micron\ fluxes vs. fluxes obtained in this
  paper (with new 3.6 \micron-based apertures) for galaxies in common. 
  Higher fluxes ($\gtrsim60$~mJy) fall close to the one-to-one line. 
  The 7 galaxies for which we measure fluxes more than $2\times$ larger 
  than the J07 fluxes have either larger apertures in our method or 
  contaminating sources (indicated with green circles, while the corrected 
  fluxes are marked with blue squares). We are using the values obtained 
  with our 3.6 \micron-defined Kron apertures for the remainder of the work. 
  Bottom: Residuals for top panel plot.} 
\label{compare}
\end{center}
\end{figure}

T10 used the MIPS photometry results from J07. In Figure
\ref{compare}, we compare the 24 \micron\ fluxes obtained in the
present work to the 24 \micron\ fluxes obtained by J07. The two
methods of defining apertures give consistent values at higher fluxes
($\ge 60$~mJy), and show more scatter ($\pm$50 per cent) at lower flux
values.  We obtain noticeably larger (factor of a few) fluxes for 7
out of 40 galaxies. After visual inspection of the discrepant sources,
we find that for three out of the 7 the apertures we have defined for
these galaxies are contaminated by other sources (e.g., foreground
stars or background galaxies).  For these galaxies (HCG 19C, HCG 22A,
and HCG 22B), exclusion regions were defined to remove the
contaminating flux. For the remaining 4 out of these 7 galaxies (HCG
31Q, HCG 42D, HCG 48C, and HCG 48D), our apertures are larger than the
apertures used by T10, and so include more light. All other galaxies
in our sample were also visually inspected for contaminating sources,
and none were found.

\begin{figure}
\begin{center}
\includegraphics[width = \columnwidth]{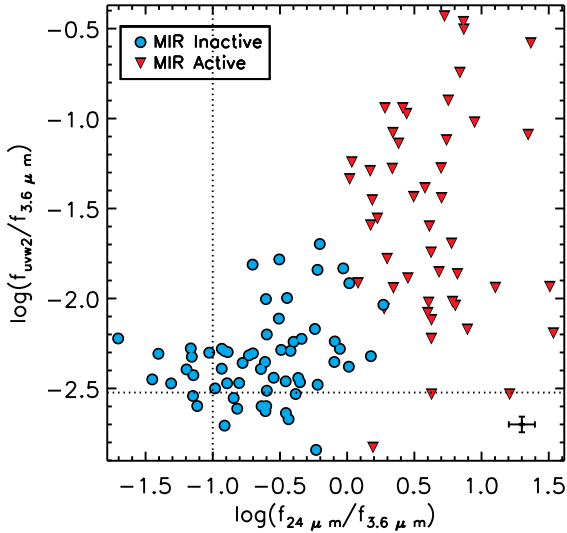}
\caption{ The $uvw2$ versus 24 \micron\ emission with both axes
  normalized to the 3.6 \micron\ fluxes. The data points are
  colour-coded according to their MIR colours. MIR-inactive galaxies
  are quiescent, and MIR-active galaxies are actively
  star-forming. The dashed horizontal line corresponds to
  UV-correction factor, $\alpha_{uvw2} = 0.003$ and the dashed
  vertical line corresponds to MIR-correction factor, $\alpha_{24} =
  0.1$.  The region above and to the right of the dotted lines
  contains galaxies whose UV and MIR light is dominated by star
  formation. Galaxies outside of this region likely have their UV
  and/or MIR light dominated by emission from old stars.}
\label{check}
\end{center}
\end{figure}

In quiescent galaxies, populations of evolved stars contribute
measurable emission in the UV (from horizontal branch stars) and at 24
\micron\ (from asymptotic giant branch stars;
\citealt{kaviraj07}). This can be seen in Figure \ref{check} where we
plot the level of $uvw2$ versus 24 \micron\ fluxes as fractions of the
3.6 \micron\ flux. Galaxies found in the region above and to the right
of the dotted lines have both UV and MIR emission dominated by star
formation, and we see that most of the MIR-active (actively
star-forming) galaxies lie in this region. Objects in the lower right
region are possibly reddened (and thus depressed UV flux) whereas the
UV and MIR light of objects in the lower left region is dominated by
old stellar populations.  To account for this, we correct the $uvw2$
and 24 \micron\ fluxes for contamination by emission from old stellar
populations before converting them to star formation rates, following
the method of \citet{ford13}. The authors used the following equations
to perform this correction

\begin{equation}
I_{\rm{FUV,SF}} = I_{\rm{FUV}} - \alpha_{\rm{FUV}}I_{\rm{3.6}}
\end{equation}

\begin{equation}
I_{\rm{24,SF}} = I_{\rm{24}} - \alpha_{\rm{24}}I_{\rm{3.6}}
\end{equation}

\noindent
where $I$ represents intensity. Additionally, $\alpha_{\rm{FUV}} =
I_{\rm{FUV}}/ I_{\rm{3.6}} = 0.003$ and $\alpha_{\rm{24}} =
I_{\rm{24}}/I_{\rm{3.6}} = 0.1$. These values were obtained by
\citet{leroy08} by looking at the ratio of fluxes in elliptical
galaxies. The $uvw2$ filter is more similar to GALEX NUV (used by
\citealt{leroy08}) than it is to FUV, and so we examine whether the
$\alpha_ {\rm{FUV}}$ parameter may have a systematic offset. In
\citet[][see their Fig. 1]{smith12}, we see that the GALEX $FUV-NUV$
colour distribution (where FUV and NUV are in magnitudes) of massive
quiescent galaxies with $r$-band luminosities of
$\gtrsim10^{9.5}$~$\mathrm{L_{r}}/\mathrm{L_{\odot}}$ is flat (with
significant scatter). Since most galaxies in our sample fall within
that range, we choose to set $\alpha_{uvw2} = \alpha_{\rm{FUV}}$.

\subsection{Stellar Masses}

\begin{figure}
\begin{center}
\includegraphics[width = \columnwidth]{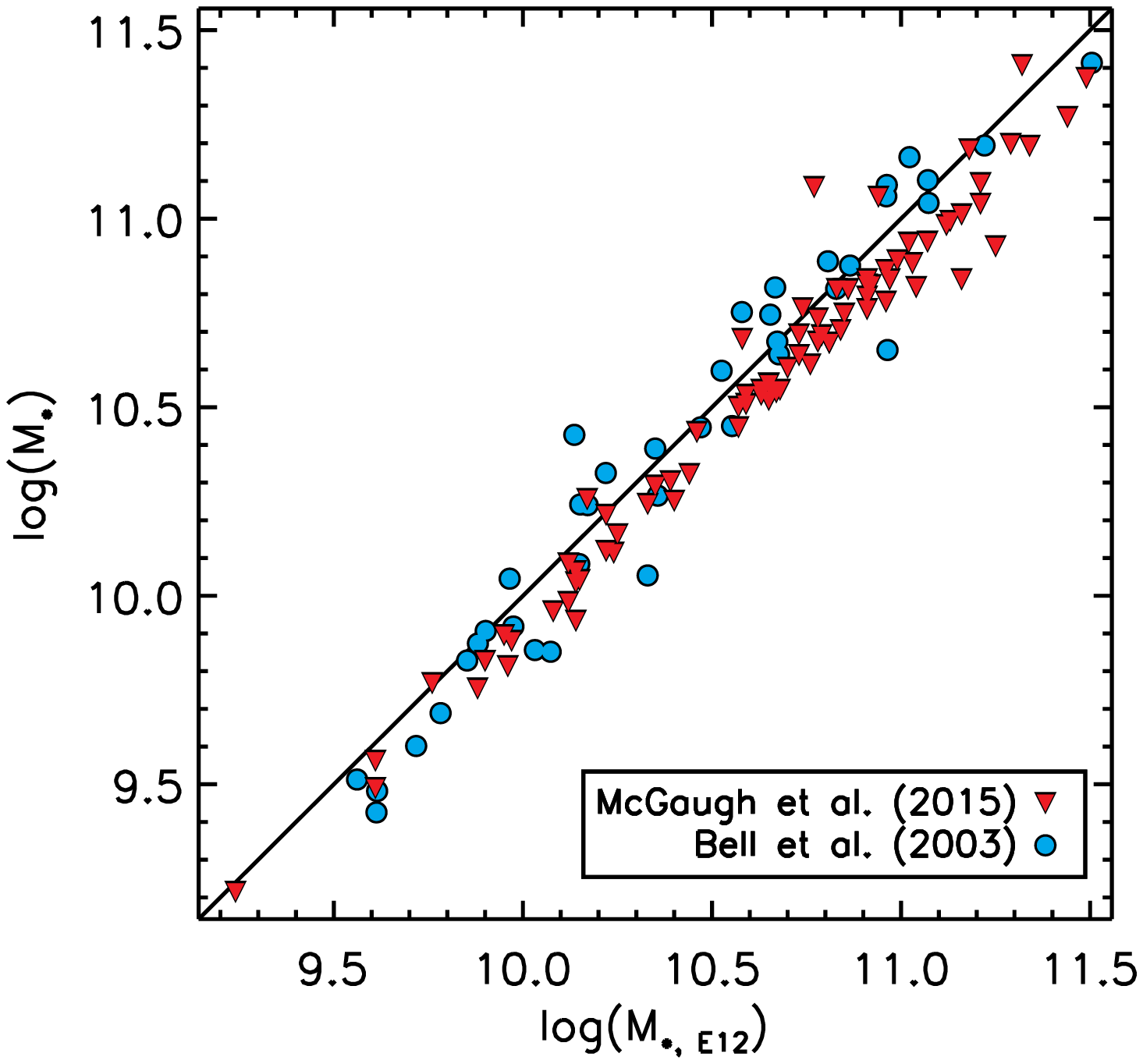}
\caption{Log of the stellar masses derived following \citet{bell03}
  and \citet{mcgaugh15} vs. the log of the stellar masses derived from
  the relation of \citet{eskew12} used in this paper. The one-to-one
  line is overplotted. The stellar masses derived from the relation of
  \citet{mcgaugh15} are very well correlated to the stellar masses
  derived in this work and they fall below the one-to-one line by a
  median of about 20 per cent. The stellar masses derived from the
  method of \citet{bell03} are also well correlated to the stellar
  masses derived in this work, and show only slight scatter about the
  one-to-one line.}
\label{sm}
\end{center}
\end{figure}

Stellar masses were calculated from the IRAC 3.6 and 4.5 \micron\
photometry according to the prescription of \citet{eskew12}:

\begin{equation}
\label{mass_stellar}
M_{*} = 10^{5.65}F_{3.6}^{2.85}F_{4.5}^{-1.85}\left( \frac{D}{0.05} \right)^{2} (\mathrm{M_{\odot}})
\end{equation}

\noindent
where $D$ is the luminosity distance of each CG given in
Table~\ref{sample}. In brief, \citet{eskew12} used extinction and star
formation history maps of the Large Magellanic Cloud to calculate its
stellar mass, and calibrate the 3.6 and 4.5 \micron\ fluxes for the
purpose of stellar mass measurements. We compare this method to that
of \citet{bell03} and \citet{mcgaugh15}, and present the results in
Figure \ref{sm}. \citet{bell03} use a sample of galaxies from the Two
Micron All Sky Survey and the Sloan Digital Sky Survey to derive a
luminosity and stellar mass function, based on $K$-band
luminosities. We compare the stellar masses calculated according to
\citet{eskew12} with the stellar masses of \citet{tzanavaris10}, which
are calculated using the relation of \citet{bell03)}, and see that
there is good agreement between these two methods, with some slight
scatter about the one-to-one line. \citet{mcgaugh15} use a sample of
26 galaxies to constrain the slope of the Baryonic Tully-Fisher
relation and use this to derive $K$-band and 3.6
\micron\ mass-to-light ratios that are independent of the initial mass
function. Their sample goes down to a mass of 3 $\times$ 10$^{7}$
M$_{\odot}$. We have used their 3.6 \micron\ mass-to-light ratio to
calculate stellar masses and compare to our values. Our stellar masses
agree well with those calculated according to \citet{mcgaugh15}. We
find that all of our stellar masses are in very good agreement with
those derived using the method of \citet{mcgaugh15} and
\citet{bell03}, and proceed with the values we calculated with the
relation of \citet{eskew12}. Note that PAH emission from starbursting
dwarfs (e.g., HCG 31F) may cause the masses to be overestimated in
these types of galaxies \citep{desjardins15}.

\subsection{Star Formation Rates}

Active star formation in galaxies can be traced directly by studying
UV emission from massive O and B stars and indirectly by studying IR
emission, from dust heated by the light from these stars. UV-based
SFRs must be corrected for intrinsic extinction in order to avoid
underestimating their true values \citep{kennicutt98, salim07},
whereas IR-based values of SFR assume that most of the UV emission is
absorbed and re-emitted in the IR \citep{calzetti07,
  rieke09}. Therefore, a complete picture of the star formation in our
sample of compact groups can be obtained by measuring the
dust-obscured component of the SFR from the MIR emission, and the
unobscured component of the SFR from the UV light. In this section, we
are following the methodology of T10, and briefly describe it here.

The $uvw2$ filter has an effective wavelength of 2030 \AA, which is
approximately in the middle of the UV \mbox{1500--2800 \AA}
range. This therefore provides a ``mean" estimate of the UV emission
properties of CGs in our sample. Hence, for this study, the $uvw2$
filter luminosities were used to determine the UV contribution to
SFR$_{\rm TOTAL}$. The relation between luminosity and SFR$_{\rm UV}$
is given by equation~\ref{sfr_uv} \citep{kennicutt98}. Equation
\ref{sfr_ir} \citep{rieke09} uses the calorimeter assumption (e.g.,
completely obscured star formation) calibrated to
24\micron\ luminosities; we use this relation to measure the IR
component of SFR$_{\rm TOTAL}$. Finally, the total SFR is measured by
simply adding the two components together, as in equation
\ref{sfr_tot}.

\begin{equation}
\label{sfr_uv}
{\rm SFR}_{uvw2} (M_{\odot }\, {\rm yr^{-1}}) \equiv 9.5 \times 10^{-44}\nu L_{\nu,uvw2} \ ({\rm erg\, s^{-1}})
\end{equation}

\begin{equation}
\label{sfr_ir}
{\rm SFR}_{24 \mu m} (M_{\odot }\, {\rm yr^{-1}}) \equiv 2.03 \times 10^{-43}\nu L_{\nu,24 \mu m} \ ({\rm erg\, s^{-1}})
\end{equation}

\begin{equation}
\label{sfr_tot}
{\rm SFR}_{TOTAL} \equiv {\rm SFR}_{uvw2} + {\rm SFR}_{24 \mu m}
\end{equation}

Our sample includes several early-type galaxies. \citet{davis14} use
the results of \citet{hao11} to combine 22 \micron\ WISE data and
GALEX FUV data to derive SFRs for early-type galaxies in the
ATLAS$^{3\mathrm{D}}$ sample. We note that the \citet{hao11}
near-UV+25 \micron\ calibration (SFR [M$_{\odot}$ yr$^{-1}$] $= 1.1
\times 10^{-43}L_{\rm 2000{\textup{\AA}}}$ [erg s$^{-1}$] $+ 2.4
\times 10^{-43}L_{25\mu m}$ [erg s$^{-1}$]) corresponds well to our
version. Although \citet{davis14} use FUV instead, as was discussed in
section 2.5, the GALEX $NUV-FUV$ distribution for galaxies with r-band
luminosities $\gtrsim$ 10$^{9.5}$~$\mathrm{L_{r}}/\mathrm{L_{\odot}}$
is flat. Since most CG galaxies in our sample satisfy this condition,
we do not expect that the method of \citet{davis14} for calculating
SFRs will yield results that are substantially different from the ones
presented in this work.

Specific star formation rate (SSFR) values were also calculated in
order to probe the star formation history of the galaxies in our
sample, in addition to studying current star formation. SSFR is given
by:

\begin{equation}
{\rm SSFR} \equiv \frac{{\rm SFR}}{M_{*}}
\end{equation}

where the stellar mass values were obtained from equation~\ref{mass_stellar}.

\subsection{Comparison to \citet{bitsakis14}}
For the 93 galaxies that we have in common, we have compared our SFR,
stellar mass, and SSFR values to those of \citet{bitsakis14}. We find
that the SFR and stellar mass results derived in this work are
systematically higher than those of \citet{bitsakis14}(median of 132
per cent and 2 per cent higher respectively). The SSFRs are observed
to be scattered about the one-to-one line. In order to understand the
offsets, we compare photometry, and examine the methodologies and
assumptions of both studies.

\subsubsection{Photometry}
\begin{figure}
\begin{center}
\includegraphics[width = \columnwidth]{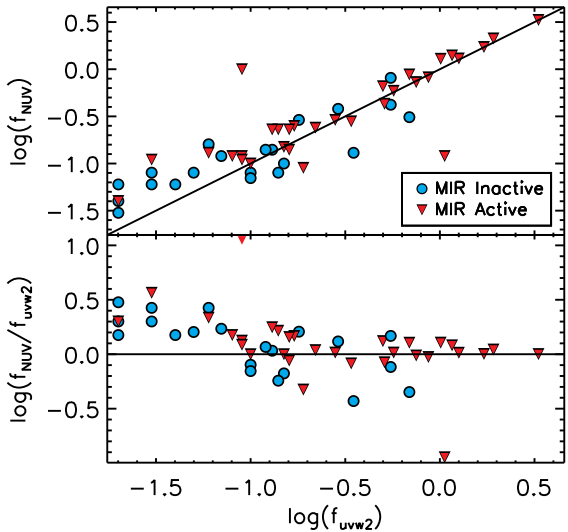}
\caption{Top: GALEX NUV fluxes of \citet{bitsakis14} plotted against
  the \textit{Swift} $uvw2$ fluxes of this work, separated according
  to MIR activity. In general, the NUV fluxes are higher than the
  $uvw2$ fluxes for fainter galaxies, however there does not seem to
  exist any relation to the MIR activity. Bottom: Residuals for the
  top panel.}
\label{uv_comp}
\end{center}
\end{figure}

There are differences in photometry that could partially account for
the systematic discrepancies. \citet{bitsakis14} used 3.6
\micron\ data to define apertures, much like we did. These apertures
were used across all of their data, however none of their data was
convolved to the resolution of the largest beam (of {\em Herschel}
SPIRE 500 \micron).  Instead, if the beam size was larger than the 3.6
\micron-defined aperture they used the beam size to extract the
photometry (T. Bitsakis, 2015, priv.  comm.).  Further, they also use
slightly different luminosity distances from those used in our work.
To mitigate this effect, we have compared our measured UV and MIR
fluxes directly to those of \citet{bitsakis11}. The IR fluxes are in
good agreement, while the UV fluxes in Figure \ref{uv_comp} show some
scatter about the one-to-one line. We define offsets as the fractional
difference between the linear fluxes we derive ($f_{\rm L15}$) and the
fluxes of \citet{bitsakis11} ($f_{\rm B11}$))

\begin{equation}
\frac{f_{\rm L15} - f_{\rm B11}}{f_{\rm L15}}
\end{equation}

The UV flux offsets have a mean of $-51$ per cent, a median value of
$-21$ per cent and a standard deviation of 150 per cent, while the IR
flux offsets have a mean, median and standard deviation of $-2$ per
cent, 5 per cent, and 72 per cent respectively. Thus, the GALEX NUV
fluxes of \citet{bitsakis11} are in general higher than the UV fluxes
we derive and this can be seen from Figure \ref{uv_comp}, while the IR
fluxes are in much better agreement.

\citet{kaviraj07} showed that UV fluxes can become contaminated by
post-starburst stellar populations. While this possibility cannot be
excluded, we note that there are no extreme UV outliers in the UV
vs. MIR luminosity plots in Figure \ref{uv_ir}. It is thus unlikely
that this is a major concern for our galaxies.

Emission from hot dust around AGN can contaminate the 24 \micron\ data
\citep{sanders89}; however \citet{tzanavaris14} showed that out of 37
HCG galaxies, only two are strong AGN while the remaining have X-ray
nuclei consistent with low-luminosity AGN or a circumnuclear X-ray
binary population. In addition, the MIR fluxes derived in this work
agree well with those derived by \citet{bitsakis14}, therefore it is
unlikely that this might be the cause for the discrepancy between this
work and the work of \citet{bitsakis14}.

\subsubsection{Methodology}
To assess the contribution of differences in methodology between this
work and the work of \citet{bitsakis14}, we use the NUV and 24
\micron\ fluxes from \citet{bitsakis11} to derive SFRs using our
methodology. By doing this, the SFRs calculated using our method and
the photometry of \citet{bitsakis11} show a median increase of
$\approx$90 per cent, relative to the SFRs reported in
\citet{bitsakis14}.

The greatest differences are seen at the low end of the SFR range of
values (quiescent galaxies); our method estimates higher SFRs for
quiescent galaxies than does the method of \citet{bitsakis14}. It is
important to note that galaxies which \citet{bitsakis14} find to be
low star formation rate galaxies are also identified as such using our
method. For example, \citet{bitsakis14} report a SFR of 0.01
$M_{\odot}$ yr$^{-1}$ for HCG 15C, while we find a SFR of 0.12
$M_{\odot}$ yr$^{-1}$. For more actively star forming galaxies, the
increases in SFRs (our values versus theirs) are much smaller. This
accounted for some of the systematic discrepancy. There is a remaining
median difference of $\approx$43 per cent. We attribute the rest to
differences in methodology, which we further discuss.

\citet{bitsakis14} use SED-fitting of the observed UV to
sub-millimeter (152.8 nm to 500 \micron) luminosities to estimate the
star formation rates and stellar masses of individual galaxies. They
use the MAGPHYS code \citep{daCunha08} to interpret their SEDs and
obtain these physical parameters. MAGPHYS is a stellar population
synthesis model which uses the Galactic disk initial mass function
(IMF) of \citet{chabrier03}. The version of MAGPHYS used by
\citet{bitsakis14} employs the \citet{bruzual03} stellar-population
synthesis models because the libraries of Charlot \& Bruzual (2007)
overestimate the importance of the thermally pulsing asymptotic giant
branch (TP-AGB), and therefore inaccurately overpredict the near-IR
luminosities in the stellar evolution models, leading to an
underestimate of stellar masses \citep{zibetti12}.

In comparison, the calibrations of \citet{kennicutt98} and
\citet{rieke09} that we use to calculate SFRs assume a Salpeter IMF
and do not depend on population synthesis models. The model of
\citet{eskew12} we use to determine stellar masses also does not rely
on population synthesis models.  Rather, the authors performed star
counts in the Large Magellanic Cloud with NIR data, and used this to
calibrate a conversion between 3.6 and 4.5 \micron\ fluxes, and
stellar mass. This approach to determining stellar mass still has some
dependence on the IMF and on stellar evolution models, however, it
does not depend on the star formation history and therefore minimizes
the effects of rare stages of stellar evolution such as the TP-AGB
phase. Their model favours bottom-heavy IMFs such as that of
\citet{salpeter55}, and disfavours bottom-light IMFs such as that of
\citet{chabrier03}.

As an independent check, we have derived stellar masses using the 3.6
\micron\ mass-to-light ratio of \citet{mcgaugh15} and have found that
they correlate very well with the stellar masses we have derived using
the relation of \citet{eskew12}. The method of \citet{mcgaugh15} is
independent of the IMF and the authors find a 3.6
\micron\ mass-to-light ratio of 0.45 M$_{\odot}$/L$_{\odot}$, compared
to 0.5 M$_{\odot}$/L$_{\odot}$ found by \citet{eskew12}.

\subsubsection{Conclusion}
After performing this comparison, we are confident that we have ruled
out any unknown systematics in our treatment that leads to the mostly
small discrepancies with \citet{bitsakis11}. The most important point
to make however, is that despite the differences in derived values,
our results are not affected qualitatively. The bimodality we observe
in SSFRs remains intact regardless of what values we choose to use.
Therefore, we now proceed with our analysis using our derived values.
 
\section{Results and Discussion}
\subsection{UV Results}

\begin{figure}
\begin{center}
\includegraphics[width = \columnwidth]{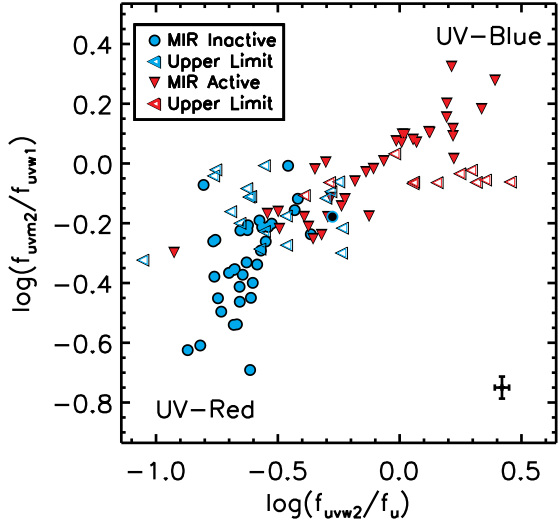}
\caption{UV colour-space distribution, defined in terms of flux
  ratios, of 101 galaxies from our sample of 192 CG galaxies. The
  hollow data points are those which suffer from coincidence loss in
  the $u$ filter and are therefore upper limits (along the horizontal
  axis). MIR-active galaxies are found to be UV-blue, as expected if
  they are actively star-forming galaxies, while the MIR-inactive
  galaxies are found to generally be UV-red, consistent with being
  quiescent galaxies. The median error bars are presented in the
  bottom right corner.}
\label{UV_Colors}
\end{center}
\end{figure}

In Figure \ref{UV_Colors}, we have plotted the colour-space
distribution of 101 galaxies for which we have data in all four UV
filters. We do this to investigate whether the UV colours can
independently tell us something about the status of star formation in
the CG environment.  We have indicated in red triangles galaxies which
are MIR-active (i.e., actively star-forming) and indicated in blue
circles galaxies which are MIR-inactive (i.e., quiescent) according to
the criteria of \citet{walker12}. The majority of MIR-active galaxies
are also UV-blue, and the MIR-quiescent galaxies are UV-red as
expected if ongoing star formation contributes UV-light from young,
massive stars as well as dust and PAH emission. There is some overlap
between these two populations in this particular colour-space without
a gap or canyon; the distribution is continuous. The data (especially
where there is no coincidence loss) are well correlated.  The Spearman
test supports this, returning a Spearman rank coefficient of 0.87 and
a significance of $5.9\times10^{-32}$, indicating a significant
correlation between the two UV colours. If we remove the data that
suffer from coincidence loss (the unfilled points), we find a Spearman
Rank coefficient of 0.89 with a significance of $8.6\times10^{-25}$.

We notice in this colour-space that there is a sort of plateau in
$\log(F_{\rm uvm2}/F_{\rm uvw1})$ of 8 MIR-active galaxies: HCG 2A,
HCG 2B, HCG 22C, HCG 31ACE, HCG 31G, RSCG 31B, RSCG 44C and RSCG
66A. These are all upper limits as they are saturated in the $u$
filter, and thus have only lower limits to their $F_{\rm u}$
values. Since the upper limits are in the x-axis, the true
$\log(F_{\rm uvw2}/F_{u})$ values are to the left of the open data
points.  There is also one MIR-active galaxy which appears to be
UV-red. This is the edge-on spiral galaxy HCG 37B, and therefore
likely suffers from significant reddening in the UV.

\subsection{MIR and UV Comparisons}

\begin{figure*}
\begin{center}
\includegraphics[width = \textwidth]{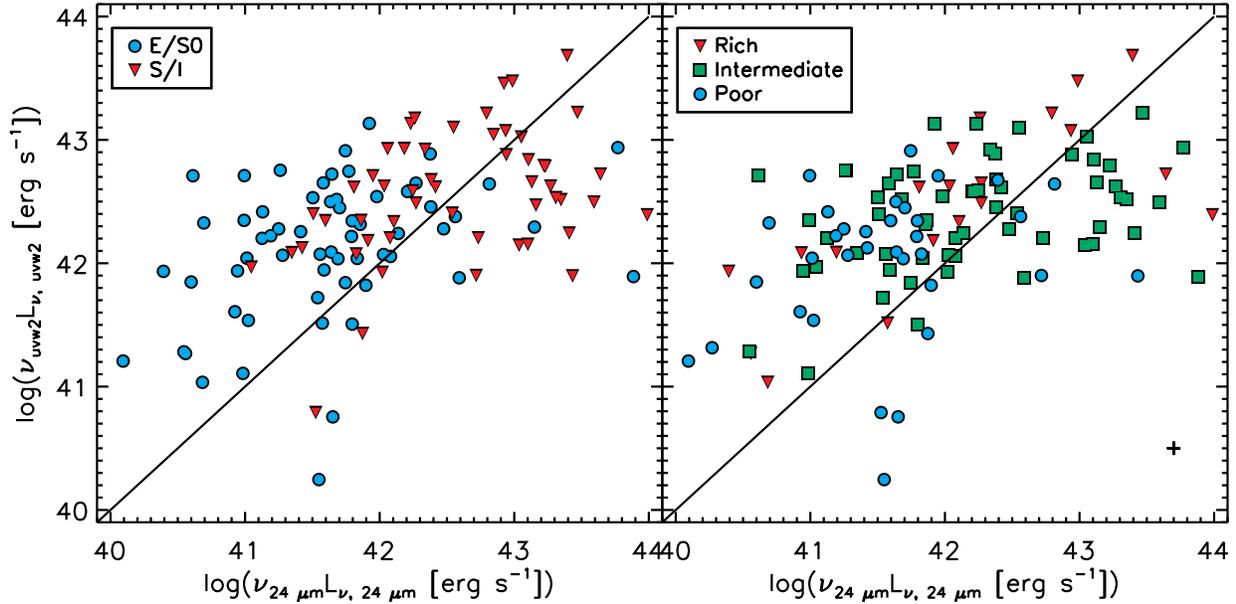}
\caption{log($\nu_{uvw2}L_{\nu,uvw2}$) vs. log($\nu_{\rm 24 \mu
    m}L_{\nu, 24 \mu m}$) for 114 CG galaxies. Left: Galaxy symbols
  indicate E/S0 (blue circles) and Sp/Irr (red triangles)
  morphologies. All CG galaxies are in general more luminous at UV
  wavelengths, than MIR. Right: Galaxy symbols indicate parent group
  H~{\sc i} type. Galaxies in gas-rich groups are indicated with
  squares, intermediate groups with circles, and poor groups with
  triangles. These parent group H~{\sc i} types are defined according
  to dynamical mass, as found in J07. Galaxies in H~{\sc i}-rich
  groups tend to be more luminous than those in H~{\sc i}-poor groups,
  while galaxies residing in groups of intermediate H~{\sc i} content
  span the range of luminosities. For reference, a one-to-one line is
  overplotted in both panels. Median error bars are presented in the
  bottom right corner of the right panel.}
\label{uv_ir}
\end{center}
\end{figure*}

We have plotted the $uvw2$ and 24 \micron\ luminosities for individual
galaxies in Figure \ref{uv_ir} and have indicated the galaxies
according to morphology (E/S0 or S/Irr) and according to the
parent-group H~{\sc i} richness for each galaxy. The luminosities are
mildly correlated (Spearman Rank coefficient of 0.58 with a
significance of $1.9\times10^{-11}$) but with considerable scatter
that increases at $\sim$ 10$^{42}$ erg s$^{-1}$.  When coded by
morphology, it becomes clear that E/S0 galaxies are typically less
luminous at both UV and MIR wavelengths, even though they are also
typically more massive.  This is expected since we would expect to see
more star formation in S/Irr galaxies. We also notice that the
galaxies in our sample are generally more luminous in the UV rather
than the MIR.  This implies that there are not significant amounts of
embedded star formation only detectable at 24 \micron.

However, there is a population of galaxies which fall below the
one-to-one line indicating that they have stronger MIR emission
compared to the UV.  The majority of these galaxies are S/Irr, and so
they may have significant dust along the line of sight that depresses
the UV emission relative to the MIR.  From visual inspection of the
\textit{Swift} $u$ images, we find that of the 29 galaxies which fall
to the right of the one-to-one line, 7 have E/S0 morphologies, 5
galaxies appear to be (near) face-on spirals, and approximately half
(17) are high inclination (close to edge-on) spirals.  None of the
MIR-brighter E/S0 galaxies (HCGs 2B, 37E, 90B and RSCGs 4B, 4C, and
17C) are known AGN according to
NED\footnote{\url{http://ned.ipac.caltech.edu/}}. Furthermore, most of
the galaxies which fall below the one-to-one line reside in groups of
intermediate H~{\sc i} gas-richness. This could be due to there not
being a large number of H~{\sc i}-rich groups, and H~{ \sc i}-poor
groups do not contain many spiral galaxies. The overall group type
does not segregate galaxies in this parameter space, though we see a
general trend that galaxies in H~{\sc i}-poor groups have lower UV and
MIR luminosities, and correspond to E/S0 type galaxies. Galaxies in
H~{\sc i}- rich groups generally have higher UV and MIR luminosities
and correspond more to S/Irr galaxies. Galaxies in groups of
intermediate H~{\sc i}-richness span the range of UV and MIR
luminosities.

\begin{figure}
\begin{center}
\includegraphics[width = \columnwidth]{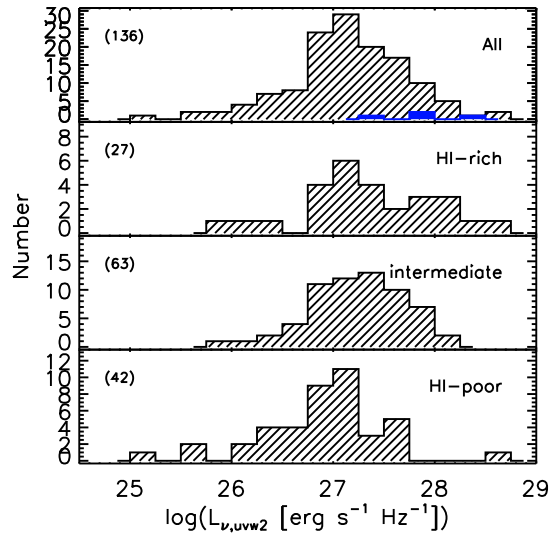}
\caption{Histogram of the $uvw2$ monochromatic luminosities for the
  full sample of CGs. The distribution for the entire sample is shown,
  as well as subsamples defined by the group H~{\sc i} content. Data
  points HCG 71A, HCG 71B, HCG 71C and RSCG 66A have an undetermined
  H~{\sc i} group type and are therefore plotted in solid blue in the
  top panel. The number of galaxies in each panel is given in
  parentheses in its upper left corner.}
\label{L_nu}
\end{center}
\end{figure}

In Figure \ref{L_nu}, we have plotted the distribution of $uvw2$
luminosities for the full sample of CGs which have available data in
this filter, as well as subsamples based on the total H~{\sc i}
content of the group. Using the definition of J07, gas richness is
defined as: H~{\sc i} rich groups have $\log(M_{\rm HI})/\log(M_{\rm
  dyn}) \ge 0.9$, H~{\sc i} intermediate groups have $0.9 >
\log(M_{\rm HI})/\log_{10}(M_{dyn}) \ge 0.8$ and H~{\sc I} poor groups
have $\log_{10}(M_{H~{\sc I}})/\log_{10}(M_{dyn}) < 0.8$.  In contrast
to what was found by T10 (see their Figure 10), with this larger
sample, we see considerable overlap between the $uvw2$ luminosities of
galaxies in the three levels of H~{\sc i}-richness. We have, however,
performed a Mann Whitney U test (RS\_TEST in IDL) on the three
subsamples defined by H~{\sc i} group type and have found the
following probabilities that the subsamples have the same median
distribution: 0.221, 0.004 and 0.001, when comparing the galaxies in
rich and intermediate groups, rich and poor groups, and intermediate
and poor groups. Probabilities that are lower than the 0.05
significance level indicate that the two samples being tested do not
have the same median distribution. This suggests that the H~{\sc
  i}-rich and intermediate groups share the same median distribution,
while the H~{\sc i}-rich and poor, and H~{\sc i}-intermediate and poor
groups are different. We also performed the two-sided
Kolmogorov-Smirnov test (KS\_2SAMP in Python) and found similar
results. The $p$-values returned were 0.61, 0.04, and 0.01 when
comparing H~{\sc i}-rich and intermediate groups, rich and poor, and
intermediate and poor, respectively. The $p$-value for the H~{\sc
  i}-rich and intermediate groups is high, 61 per cent, and therefore
we can consider these two distributions to be statistically
consistent. The $p$-values for the H~{\sc i}-rich and poor, and
intermediate and poor groups, are both less than 5 per cent so we can
consider them to be statistically different. Rich, intermediate and
poor groups seem to all have more galaxies at intermediate
luminosities.  Rich groups seem to have somewhat more UV-luminous
galaxies than intermediate and poor groups, and poor groups seem to
have somewhat UV-fainter galaxies than rich and intermediate. This is
consistent with more cold gas present in rich groups enabling higher
levels of star formation.

\subsection{SFR and SSFR Results}

Our SFR$_{\rm UV}$, SFR$_{\rm IR}$ and SSFR values, along with the UV
and IR luminosities are given in Table \ref{props}. This Table also
includes stellar masses (calculated with equation \ref{mass_stellar}),
and $\alpha_{\rm IRAC}$ values of each galaxy.  The $\alpha_{\rm
  IRAC}$ parameter was defined by \citet{gallagher08} as the spectral
index of the power-law fit to the 4.5, 5.6, and 8.0
\micron\ luminosity densities, $l_{\nu}\propto\nu^{\alpha_{\rm
    IRAC}}$.  MIR-active galaxies (with red MIR spectral energy
distributions from warm dust and PAH emission) have $\alpha_{\rm IRAC}
\le 0$, whereas MIR-quiescent galaxies (with blue MIR SEDs consistent
with stellar photospheric emission) have $\alpha_{\rm IRAC}>0$. The
$\alpha_{\rm IRAC}$ values presented in Table \ref{props} were
measured by fitting a power-law to the \citet{walker12} IRAC
photometry.

\begin{table*}
\caption{UV and IR Properties of HCG and RSCG Galaxies. The full table is available online.}
\label{props}
\begin{tabular}{lcrrrrrrcccc}

\hline
\hline
CG & Galaxy & \multicolumn{5}{c}{$\log$ L$_{\nu}$ (erg s$^{-1}$ Hz$^{-1}$)} & {\rm log M$_{*}$} & $\alpha_{\rm IRAC}$$^{\mathrm{a}}$ & SFR$_{uvw2}$ & SFR$_{24 \mu m}$ & SSFR \\
ID & Morphology & $uvw2$ & $uvm2$ & $uvw1$ & $u$ & 24 $\mu$m & (M$_{\odot}$) &   & (M$_{\odot}$ yr$^{-1}$) & (M$_{\odot}$ yr$^{-1}$) & (10$^{-11}$ yr$^{-1}$) \\ \hline
\hline
HCG 2A	& S & 28.05 & 28.27 & 28.24 & 28.41 & 29.70 & 10.35 & -2.68 & 2.523 & 1.335 & 17.230 \\
HCG 2B	& S & 27.62 & 27.88 & 27.83 & 28.06 & 30.13 & 10.22 & -3.63 & 0.957 & 3.645 & 27.778 \\
HCG 2C	& S & 27.45 & 27.71 & 27.66 & 27.95 & 28.94 & 10.15 & -2.38 & 0.659 & 0.232 & 6.322 \\
HCG 4A	& S & $\cdots$ & 28.63 & 28.68 & 28.92 & 30.78 & 11.18 & -3.47 & $\cdots$ & 16.348 & $\cdots$ \\
HCG 4B  & S & $\cdots$ & 27.77 & 27.83 & 27.13 & 29.27 & 10.40 & -2.64 & $\cdots$ & 0.496 & $\cdots$ \\
HCG 4D  & E & $\cdots$ & 27.58 & 27.66 & 27.92 & 29.65 & 10.33 & -3.33 & $\cdots$ & 1.190 & $\cdots$ \\
HCG 7A	& S & 27.49 & 27.51 & 27.77 & 28.25 & 30.03 & 10.96 & -2.37 & 0.435 & 2.894 & 3.631 \\
HCG 7C	& S & 27.93 & 27.96 & 28.03 & 28.34 & 29.45 & 10.65 & -2.45 & 1.203 & 0.758 & 4.352 \\
HCG 7D  & S0 & 27.35 & 27.40 & 27.46 & 27.83 & 28.58 & 10.14 & -2.41 & 0.352 & 0.103 & 3.043 \\
\hline

\multicolumn{12}{p{\textwidth}}{$^{\mathrm{a}}$\citet{gallagher08} defines the $\alpha_{\mathrm{IRAC}}$ parameter. \citet{desjardins14} describes the method of SED fitting from which the values presented here were calculated.
}

\end{tabular}
\end{table*}

\begin{figure}
\begin{center}
\includegraphics[width = \columnwidth]{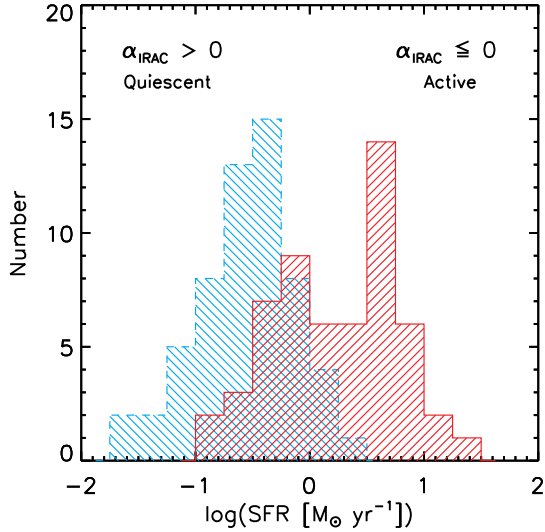}
\caption{SFR distribution for 114 CG galaxies. The sample is separated
  into two subsamples; MIR-active ($\alpha_{\rm IRAC} \le 0$; red
  hashed histogram) and MIR-quiescent ($\alpha_{\rm IRAC} > 0$; blue
  hashed histogram). Though the medians of the two populations are
  offset, the distributions have significant overlap, and the total
  distribution of SFRs is continuous.}
\label{SFR_HIST}
\end{center}
\end{figure}

Figure \ref{SFR_HIST} shows a histogram of the SFRs separated into two
subsamples: MIR-active and MIR-quiescent defined according to the
$\alpha_{\rm IRAC}$ values. Both the original sample of CGs studied by
T10 and the sample of 46 CGs studied in this project show a continuous
distribution of SFRs (Fig. 12 in T10 and our Fig.~\ref{SFR_HIST}). The
original sample from T10 shows a flatter distribution, whereas the
expanded sample is more peaked at around log(SFR) $=$ 0,
indicating that the most common SFR value in CGs might be $\approx$1
$M_{\odot}$ yr$^{-1}$. Despite the significant overlap between
MIR-active and MIR-inactive galaxies, we do see that quiescent
galaxies ($\alpha_{\rm IRAC} > 0$) are less star-forming and active
galaxies ($\alpha_{\rm IRAC} \le 0$) have higher star formation rates
on average.  The total distribution has a mean log(SFR) and standard
deviation of $-0.29 \pm 0.72$ $\mathrm{M}_{\odot}
~\mathrm{yr}^{-1}$. The population of galaxies with $\alpha_{\rm IRAC}
> 0$ has a mean log(SFR) and standard deviation of $-0.66 \pm 0.55$
$\mathrm{M}_{\odot} \mathrm{yr}^{-1}$ whereas the population of
galaxies with $\alpha_{\rm IRAC} \le 0$ has $0.04 \pm 0.69$
$\mathrm{M}_{\odot} \mathrm{yr}^{-1}$.

\begin{figure*}
\begin{center}
\includegraphics[width = \textwidth]{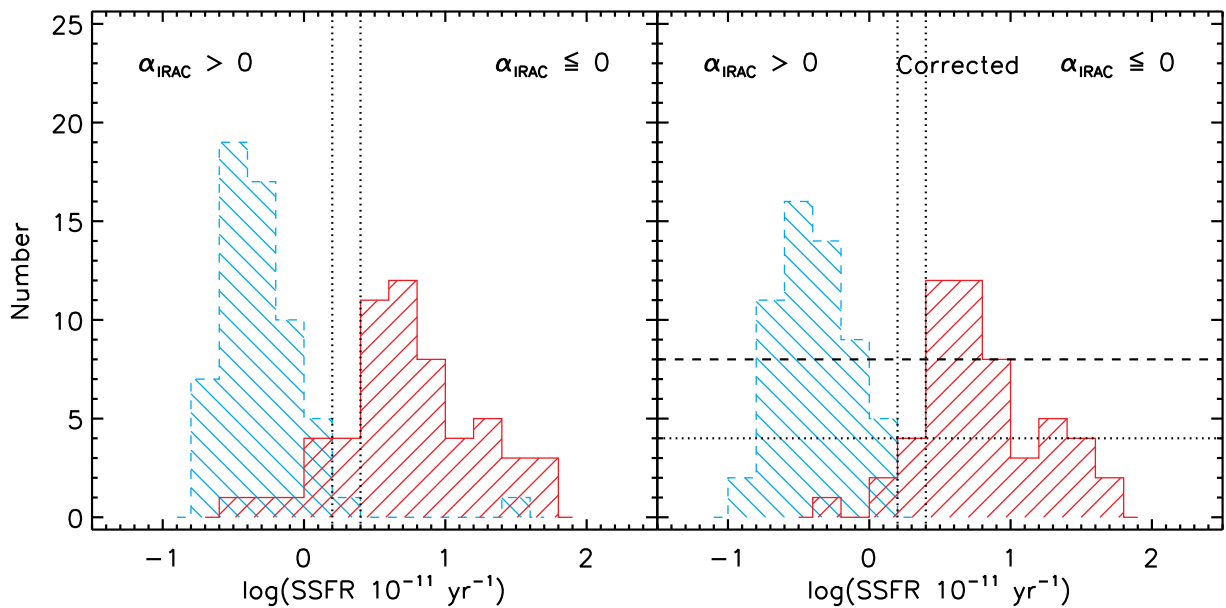}
\caption{Distribution of SSFRs for 117 CG galaxies. Left: Distribution
  of SSFRs uncorrected for old stellar populations. Right:
  Distribution of SSFRs corrected for emission from old stellar
  populations. The values in both panels are separated into two
  different populations, according to their $\alpha_{\rm{IRAC}}$
  values. Both panels show a clear bimodality in the values of
  SSFRs. The values of SSFR show even less overlap according to
  $\alpha_{\rm{IRAC}}$ values when they are corrected for emission by
  old stellar populations. The vertical lines mark the ``dip'' in the
  distribution of the total population of SSFRs, defined as half the
  median value of the histogram, where the dashed horizontal line
  indicates the median, and the dotted horizontal line marks the half
  median value.}
\label{twop_ssfr}
\end{center}
\end{figure*}

The same separation according to $\alpha_{\mathrm{IRAC}}$ does not
yield a continuous distribution for SSFRs as we can see in Figure
\ref{twop_ssfr}. In the left panel we show the distribution of SSFR
values that are not corrected for old stellar populations, whereas the
right panel shows the corrected SSFR values. Even before applying the
correction for contamination by old stellar populations, we see that
the distribution of SSFRs is bimodal. Applying the correction only
makes the dip more pronounced. The dip appears at $0.2 \le
\textrm{log} (\textrm{SSFR} [10^{-11} \mathrm {yr}^{-1}]) \le 0.4$ and
is indicated in both panels with dotted lines. This dip region is
defined as half the median value of the histogram. We performed the
Hartigan Dip Test using the R package {\tt diptest}, to check the
statistical significance of this bimodality. The Hartigan Dip Test
\citep{hartigan85} measures a distribution's departure from
unimodality by minimizing the maximum difference between an empirical
distribution and a unimodal distribution. The resulting dip statistic
(D$_{n}$) indicates that a distribution is significantly bimodal if it
is $\le 0.05$ and only marginally bimodal if it is between $0.05$ and
$0.1$. After performing this test, the resulting dip statistic,
D$_{n}$ = 0.02, indicates that the distribution is significantly
bimodal. Thus, the bimodality that was previously found by T10 is
still present even after significantly increasing the sample size,
suggesting a fast transition from active star formation to quiescence
in CG galaxies. The mean and standard deviation of the total log(SSFR)
distribution is $0.20\pm0.65$ $10^{-11} \mathrm{yr}^{-1}$. For the
population of quiescent galaxies ($\alpha_{\rm IRAC} > 0$), they are
$-0.23\pm0.42$ $10^{-11} \mathrm{yr}^{-1}$, while for the population
of actively star-forming galaxies ($\alpha_{\rm IRAC} \le 0$), they
are $0.67\pm0.51$ $10^{-11} \mathrm{yr}^{-1}$.

Since the most massive galaxies would have formed earlier in time, one
might expect that they should have lower SFRs. Due to their high
stellar masses, they will also have lower values of
SSFR. Correspondingly, we would expect galaxies with higher SFRs and
SSFRs to be lower mass galaxies because they have (consistent with
cosmic downsizing) both larger reservoirs of gas for star formation
and lower stellar masses.

\begin{figure}
\begin{center}
\includegraphics[width = \columnwidth]{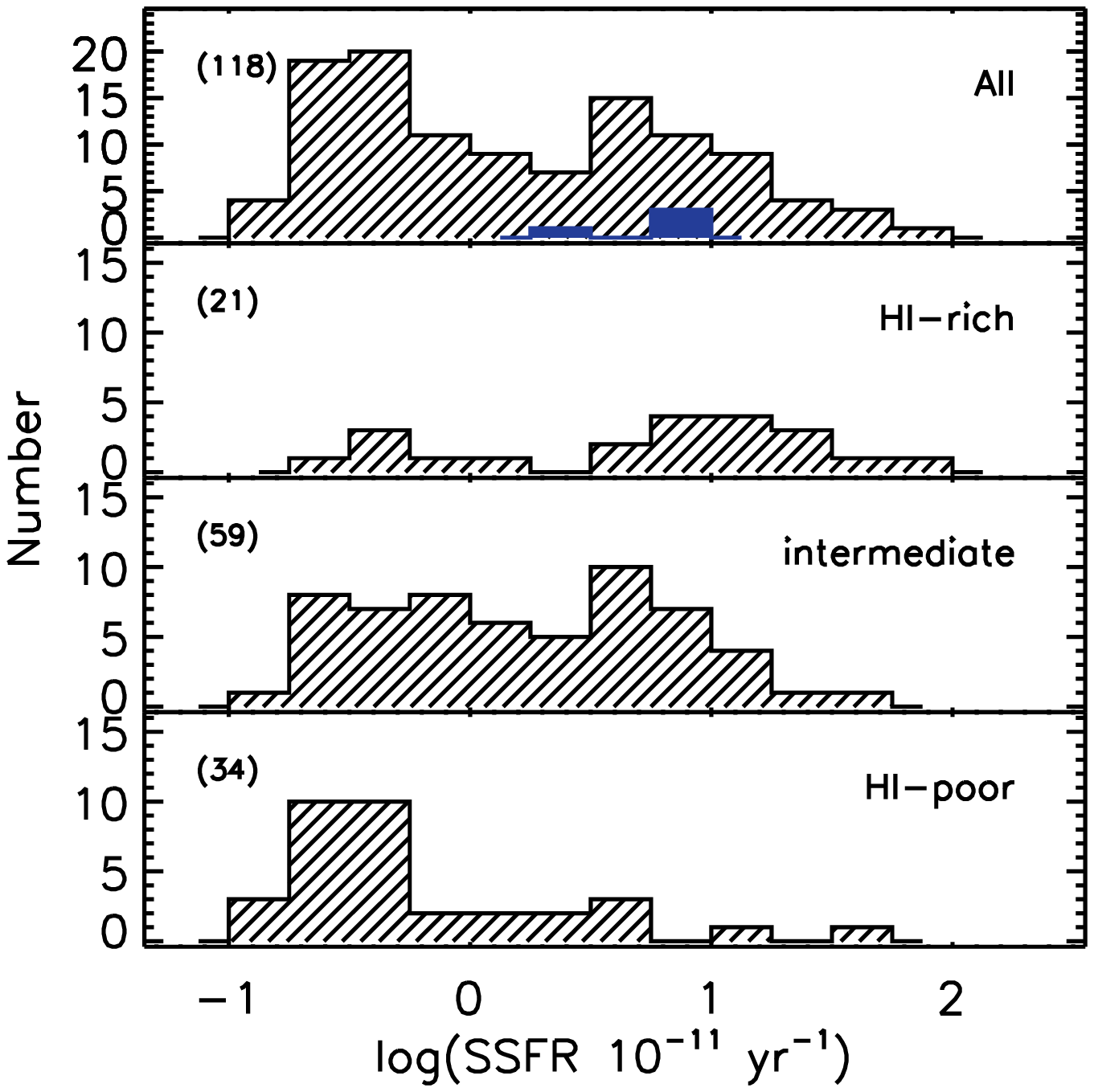}
\caption{SSFR distribution for CG galaxies. The top panel shows the
  total distribution. The H~{\sc i} group type for HCG 71A, HCG 71B,
  HCG 71C and RSCG 66A are undetermined and plotted in dark blue. The
  next three panels show subsamples determined by the group H~{\sc i}
  content. The number of galaxies in each panel is indicated in the
  left corners.}
\label{MULTI_SSFR}
\end{center}
\end{figure}

Looking at Figure \ref{MULTI_SSFR}, we notice that the SSFR values
correlate much better with H~{\sc i}-gas content of the group than did
the $uvw2$ luminosities. Rich groups tend to have higher SSFRs, groups
of intermediate H~{\sc i}-richness cover a broad range of SSFR values,
and H~{\sc i}-poor groups tend to have lower SSFRs. This is consistent
with the results from T10 (as shown in their Fig. 13).

\begin{figure}
\begin{center}
\includegraphics[width = \columnwidth]{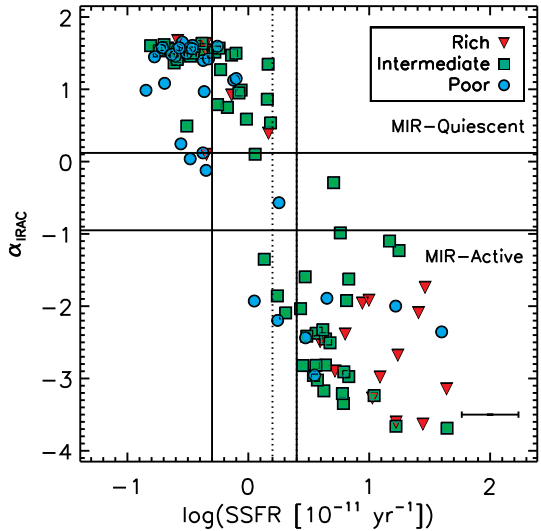}
\caption{MIR activity index, $\alpha_{\rm IRAC}$ vs. $\log(SSFR)$ for
  114 galaxies (corrected for emission from old stellar
  populations). The different plotting symbols indicate different
  parent-group gas richness. The bold lines represent the SSFR and
  $\alpha_{\rm IRAC}$ gaps observed in T10. The dashed lines show
  where we see the dip in SSFRs with the expanded sample of CGs.}
\label{alpha_irac_corr}
\end{center}
\end{figure}

In Figure \ref{alpha_irac_corr}, we have plotted the MIR activity
index, $\alpha_{\rm IRAC}$ versus SSFR. The plotting symbols
indicate the parent-group H~{\sc i} richness. As also found by T10, it
is clear from the plot that quiescent galaxies ($\alpha_{\rm IRAC} >
0$) have lower SSFRs and they mostly reside in gas-poor or
intermediate groups. We can also see that actively star-forming
galaxies ($\alpha_{\rm IRAC} \leq 0$) also have higher SSFRs and
reside mostly in gas-rich or intermediate groups. We have indicated in
dashed lines the region where we see the dip (determined from the
histogram in the right panel of Fig.~\ref{twop_ssfr}). For comparison,
we include vertical bold lines to indicate the gap region that was
previously observed by T10. The horizontal bold lines indicate gap
region in $\alpha_{\mathrm{IRAC}}$ values. Due to larger numbers, it
is perhaps not surprising that the earlier SSFR gap (T10) is now
observed as a dip, which is shown to be statistically significant.

\begin{figure}
\begin{center}
\includegraphics[width = \columnwidth]{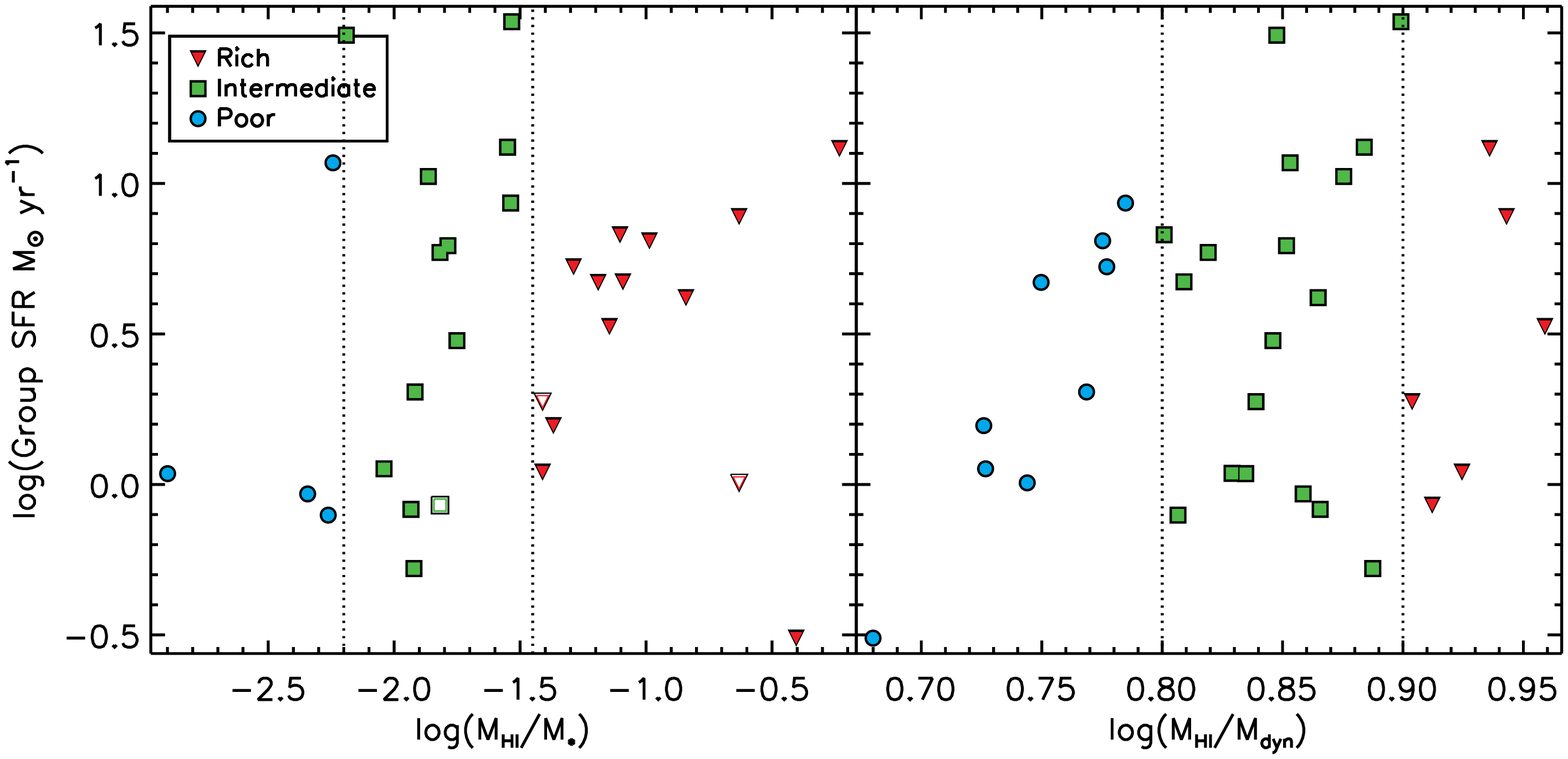}
\caption{Total group SFR vs. \mbox{$\log M_{H~{\sc I}}/\log M_{*}$}
  for every galaxy in our sample. The symbols indicate parent-group
  H~{\sc i} richness. We can see that the groups span the full range
  of SFR values regardless of their parent-group type.}
\label{sfr_mass}
\end{center}
\end{figure}

\begin{figure}
\begin{center}
\includegraphics[width = \columnwidth]{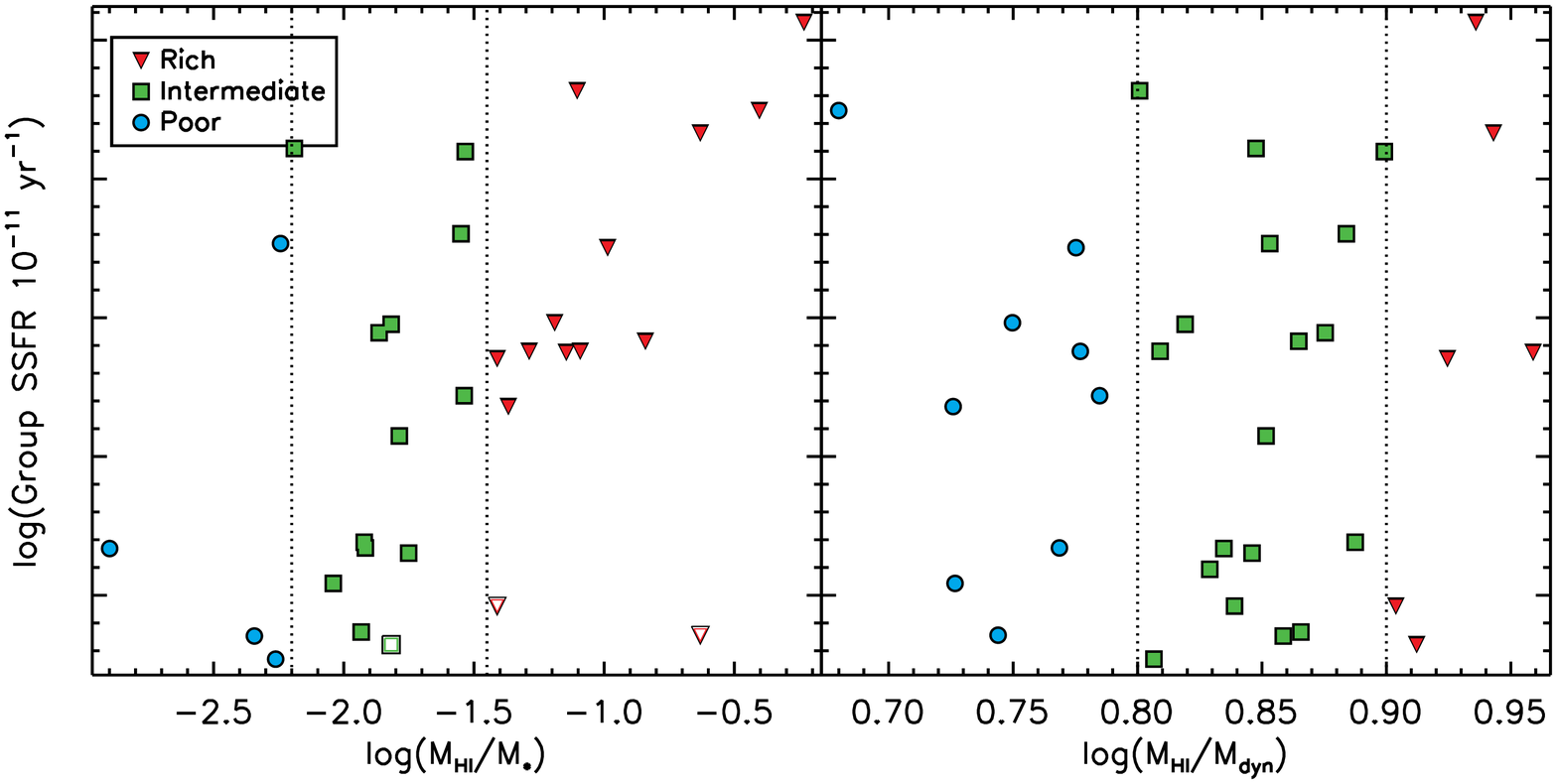}
\caption{Total group SSFR vs. \mbox{$\log M_{H~{\sc I}}/\log M_{dyn}$}
  for every group in our sample. The symbols indicate parent-group
  H~{\sc i} richness. We can see that the trend remains for H~{\sc i}
  rich groups to have higher values of SSFR, as in T10.}
\label{h1_content}
\end{center}
\end{figure}

In Figure \ref{sfr_mass}, we plot the total group SFRs versus the
H~{\sc i} content of each group and find that groups span the full
range of SFR values regardless of their group type. This again shows
that the quantity of interest is the SSFR of each group, not its
SFR. We also plot the total group SSFRs versus the H~{\sc i} content
of each group in Figure \ref{h1_content}. The left panel contains
H~{\sc i} content defined in terms of stellar mass, while the right
panel contains H~{\sc i} content defined according to dynamical mass
(J07).  When we define the H~{\sc i} content types according to
stellar mass, there is a clear trend of increasing values of SSFR with
increasing richness in parent-group H~{\sc i} content, supporting the
idea that SSFR tracks gas depletion (T10).

\begin{figure*}
\begin{center}
\includegraphics[width = \textwidth]{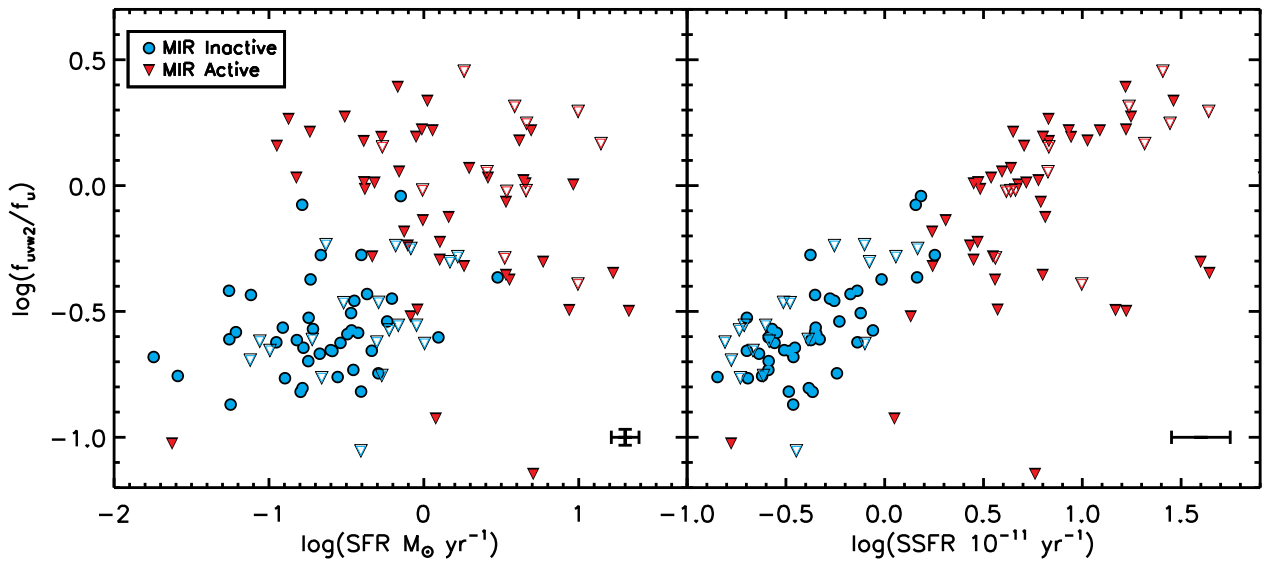}
\caption{The $f_{uvw2}/f_{u}$ UV flux ratio has been plotted versus
  SFR (left panel) and SSFR (right panel) for the 115 galaxies for
  which all the data are available. The galaxies are colour-coded as
  MIR-quiescent (blue circles) and MIR-active galaxies (red
  triangles). While the UV-flux ratio shows a scatter plot vs.  SFR,
  there is a significant trend whereby blue UV flux ratios correlate
  with increassing SSFRs. The combination of MIR-colour, UV-flux
  ratio, and SSFR measurements clearly separate the CG galaxy
  population into active and quiescent galaxies. Median error bars are
  shown in the bottom right corners of each panel.}
\label{uv_sfr}
\end{center}
\end{figure*}

Finally, we plot in Figure \ref{uv_sfr} the $\log(F_{\rm uvw2}/F_{\rm
  u})$ flux ratio versus SFR and SSFR; symbols indicate MIR-active
and MIR-inactive galaxies. When the UV flux ratio is plotted versus
SFR, there is no strong correlation. However, when we plot this UV
flux ratio against SSFR, a strong trend is evident. In this
parameter space, two populations are clearly shown: MIR-active
galaxies have higher SSFRs and tend to be UV-blue while MIR-inactive
galaxies tend to be UV-red with lower values of SSFR. There is little
overlap between the two populations. A few galaxies are discrepant:
HCG 48C, HCG 37B and HCG 92C (a strong X-ray AGN, \citealt{
  tzanavaris14}) are MIR-active but also UV-red (lower part of
plot). All three are spiral galaxies, but only HCG 48C and HCG 37B are
edge-on. Therefore, the UV-flux ratio is likely reflecting intrinsic
extinction in HCG 48C and 37B.  The X-ray AGN HCG 92C is a Seyfert 2
galaxy, and so while the direct view of the UV AGN continuum is
blocked, the AGN likely dominates the MIR emission.

We find 8 galaxies here as well which do not follow the general trend
of the data. These galaxies are all MIR-active, but have lower UV flux
ratios for their SSFRs. These galaxies are HCG 16D, HCG 40D, HCG 56B,
HCG 56D, HCG 59A, HCG 71B, RSCG 4A and RSCG 4C.

\subsubsection{Star-Forming ``Main Sequence''}
\begin{figure}
\begin{center}
\includegraphics[width = \columnwidth]{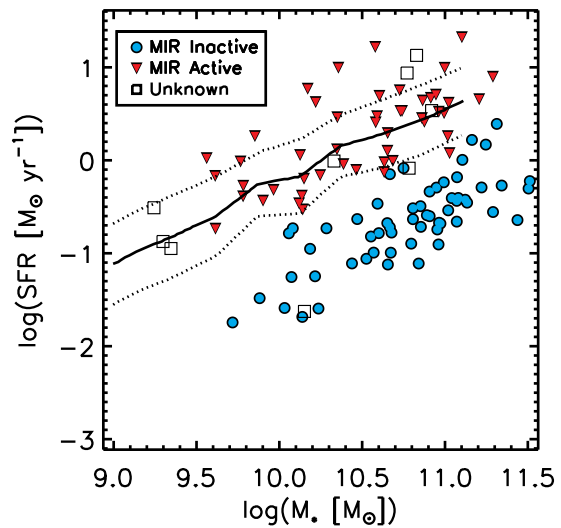}
\caption{Plot of log(SFR) vs. log($M_{\ast}$). The solid black line is
  the median main sequence line from \citet{chang15}, while the dotted
  orange lines represent the 1$\sigma$ limits. We plot the 118 CG
  galaxies for which we have SFR and M$_{*}$ measurements. We separate
  the 118 CG galaxies according to their MIR activity and find that
  all MIR inactive galaxies fall below the $-1\sigma$ limit of the
  main sequence line, while the MIR active galaxies are generally
  within $\pm1\sigma$ of the line. In cases where we do not have MIR
  activity information, we plot the data point as a black box. Based
  on their locations within this parameter space, the majority of
  these galaxies are likely to be MIR-active.}
\label{sfms}
\end{center}
\end{figure}

It has now been established that over the past 10 Gyr a correlation
exists between SFR and $M_{\ast}$ for star-forming galaxies
\citep{noeske07,elbaz07,daddi07}. \citet{wuyts11} studied large sample
of more than 800,000 galaxies at different redshifts, confirming the
correlation, albeit with a zero point that increases with redshift. We
take our SFR and stellar mass measurements, and plot our sample of CGs
in this parameter space, together with the best-fit results of
\citet{chang15}, to investigate how they compare to non-CG galaxies.
\citet{chang15} adopt a mass limit, such that all galaxies in their
sample have significant WISE 12 \micron\ emission. Their galaxies all
have redshift $< 0.6$, with the majority having redshifts $< 0.2$. We
include all of our CG galaxies regardless of mass, and show the main
sequence line, defined as the median values of 0.2 M$_{\odot}$ wide
stellar mass bins ($\pm1\sigma$) from Figure 11 of \citet{chang15} as
a comparison to our sample. This comparison is illustrated in Figure
\ref{sfms}, with galaxies separated according to MIR activity. This
compares well to the results of \citet{chang15} as all CGs plotted in
Figure \ref{sfms} fall within their range of SFRs from $-3 <
\mathrm{log(SFR [M_{\odot} yr^{-1}])} < 1$. Furthermore, we see that
MIR inactive galaxies all fall below the $-1\sigma$ limit of the main
sequence line. The MIR active galaxies however follow the main
sequence line, with most galaxies falling within the $\pm1\sigma$
limit. Of the 48 MIR-active galaxies plotted in this figure, 26 lie
above the main sequence line.

\subsection{UV-Optical Results}

\begin{figure}
\begin{center}
\includegraphics[width = \columnwidth]{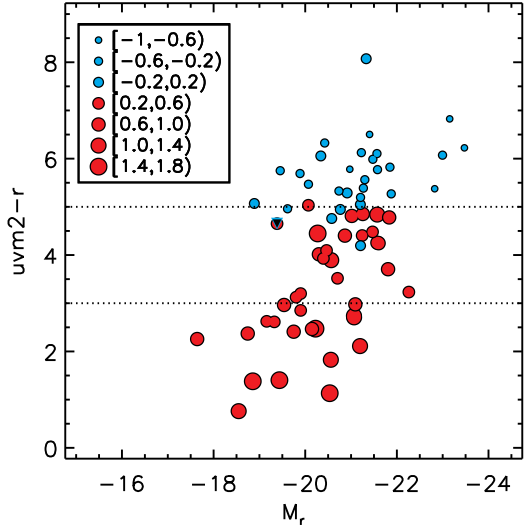}
\caption{UV-optical colour-magnitude diagram for 69 galaxies in our
  sample. The number of data points in this figure is limited to
  galaxies for which we were able to determine SSFRs. The symbol sizes
  refer to bins of SSFRs with increasing symbol size corresponding to
  increasing SSFRs, as indicated in the legend. The bins have width of
  0.4 10$^{-11} \mathrm{M_{\odot}}$. We see that all galaxies along
  the red sequence ($uvm2-r > 5$) have lower values of SSFR. The
  region of the blue cloud ($uvm2-r < 3$) contains galaxies that have
  higher SSFRs. The green valley is populated by galaxies of
  intermediate SSFR levels. No corrections have been made for internal
  extinction.}
\label{hammer_ssfr}
\end{center}
\end{figure}

Figure \ref{hammer_ssfr} is a UV-optical colour-magnitude plot. To be
consistent with earlier {\em GALEX} work, we use data from the {\em
  Swift} $uvm2$ filter which is most similar to the {\em GALEX} NUV
band.  The bottom portion of Figure \ref{hammer_ssfr} is the region of
star formating galaxies and these galaxies are accordingly
MIR-active/UV-blue, the top portion represents the red sequence where
passive or reddened galaxies reside and are accordingly
MIR-quiescent/UV-red.

Figure 12 of \citet{wyder07} shows the volume density of galaxies as a
function of $NUV-r$ for 0.5 mag wide bins. They note an
underpopulation of galaxies in the region $3 < NUV-r < 5$ mag and
refer to this as the ``green valley''. We mark this region with
horizontal dashed lines in Figure \ref{hammer_ssfr}. The black data
point in our Figure \ref{hammer_ssfr} is the galaxy HCG 37E, a MIR
canyon galaxy, as defined in \citet{walker12} and just falls in this
region of the UV-optical CMD.  Otherwise there is no evidence for a
green valley in this figure. This region is well populated with
galaxies. However, we have not corrected our UV-optical colours for
intrinsic extinction.

Comparing Figure \ref{hammer_ssfr} to Figure \ref{uv_sfr}, we see that
the UV-only colours are much cleaner than the UV-optical colour in
separating galaxies according to MIR activity. The UV-only colour may
be better at discriminating between low and high SSFRs because the
UV-colour is sensitive to SSFRs as seen in Figure
\ref{uv_sfr}. However, in addition to this, the $uvw2$ and $u$ filters
have a much narrower wavelength separation than do $uvm2$ and $r$. As
long as there is no bump at around 2200 \AA\ in the extinction curve
\citep{stecher69}, the UV-only colour should be less susceptible to
extinction.

We separate the data in Figure \ref{hammer_ssfr} according to the MIR
activity of each galaxy. We also scale the data point symbols
according to SSFR. The data are separated in bins of width equal to
0.4 $10^{-11} \textrm{yr}^{-1}$. The MIR-inactive galaxies all have
lower SSFR values and all the MIR-active galaxies have higher SSFR
values, as we would expect. It shows however, that the galaxies in the
UV-optical green valley region, which are a mixture of MIR-active and
MIR-inactive galaxies, all have intermediate SSFR values.

\begin{figure}
\begin{center}
\includegraphics[width = \columnwidth]{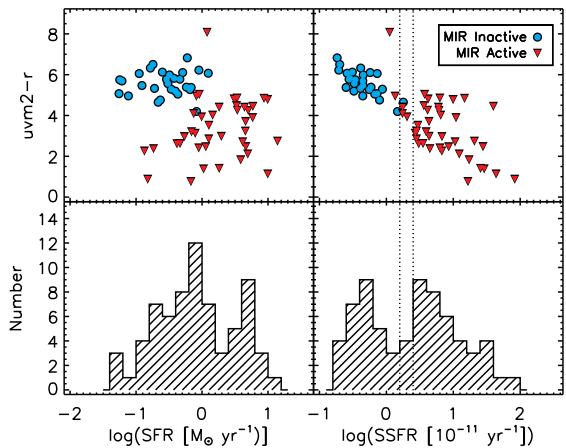}
\caption{UV-optical colour vs. SSFR and SFR for the 69 galaxies for
  which we had available data. Top left: We plot the $uvm2-r$ colour
  vs. SFR and see that the data are scattered. We colour code the
  points according to MIR activity. We see that MIR active galaxies
  are generally more star forming and have bluer $uvm2-r$ colours. MIR
  inactive galaxies are less star forming and have redder $uvm2-r$
  colours. Top right: We plot the $uvm2-r$ colour vs SSFR and
  find a much stronger correlation. Bottom left: Distribution of SFR
  values. Bottom right: Distribution of SSFRs. We use dotted lines
  again to indicate the dip region in SSFR values.}
\label{multi}
\end{center}
\end{figure}

Figure \ref{multi} plots the $uvm2-r$ colour vs the SFRs and the
SSFRs.  When we plot the $uvm2-r$ colour against the SFRs, there
appears to be no correlation. However, the data is separated cleanly
into two ``clouds'' delineated by the MIR colours. When we plot this
colour against the SSFRs, we see that the data are highly
correlated. This is not surprising since $uvm2$ is dominated by
massive, short-lived stars, thus it traces SFR. On the other hand, $r$
traces stellar mass since it is sensitive to emission from older,
longer-lived stars \citep{wyder07,martin07,thilker10}. We notice in
the top right panel that there appears to be a separate little cloud
of MIR-active galaxies at around $4 \le \mathrm{uvm2-r} \le 5$ mag.
These galaxies are HCG 7A, HCG 38A, HCG 38B, HCG 47C, HCG 56B, HCG
56D, HCG 59A, HCG 61A, HCG 71B, HCG 100A, and RSCG 4A. The scatter
observed in these MIR-active galaxies is likely related to extinction.

\newpage

\section{Summary and Conclusion}
We have presented \textit{Swift} UV ($uvw2$, $uvm2$, $uvw1$ and $u$)
and \textit{Spitzer} MIR (3.6 $\micron$, 4.5 $\micron$ and 24
$\micron$) photometry for 136, 130, 141, 146 and 169, 169, 154, CG
galaxies respectively. We have combined the UV and MIR photometry to
obtain star formation rates, stellar masses, and specific star
formation rates for 118, 169 and 118 galaxies, respectively.

Including corrections for the contribution to the UV and MIR emission
from old stellar populations. Compared to the earlier work of T10, the
sample size in this paper represents an increase by a factor of about
4.

The main results of this paper are summarized here.

\begin{itemize}

\item{We have confirmed the correlation between L$_{\nu, uvw2}$ and
  L$_{\nu, 24 \mu m}$ (Figure \ref{uv_ir}) and shown that E/S0
  galaxies are typically less luminous at UV and MIR wavelengths than
  S/Irr galaxies.  CG galaxies are however generally more luminous at
  UV wavelengths than MIR, implying that there is not much embedded
  star formation in this population. Galaxies in H~{\sc i}-rich groups
  are generally more luminous at UV and MIR wavelengths than those in
  H~{\sc i}-poor groups, while galaxies in groups of intermediate
  H~{\sc i} gas richness span the range of UV and MIR luminosities.}

\item{The UV $uvw2/u$ and $uvm2/uvw1$ flux ratios confirm our
  expectation that the MIR-active and UV-blue colours consistently
  trace young stellar populations. No gap or canyon is observed in
  this parameter space.}

\item{As seen by T10, the SFR distribution is continuous though more
  peaked than seen previously with the most likely SFR in compact
  group galaxies of $\sim 1 \rm{M_{\odot} yr^{-1}}$. Figure
  \ref{SFR_HIST} shows that the SFRs do not trace the MIR colours well
  as we see much overlap between MIR-active and MIR-quiescent
  galaxies. However, when we look at the distribution of SSFRs in
  Figure \ref{twop_ssfr}, we see that the MIR colours map onto the
  SSFRs very well. The $\alpha_{\rm IRAC}$ index separates the SSFR
  values into two almost distinct populations corresponding to
  actively star-forming and quiescent galaxies.}

\item{Figure \ref{MULTI_SSFR} shows that the SSFRs trace gas depletion
  because we see that galaxies in H~{\sc i}-rich groups tend to have
  higher SSFRs than the intermediate or poor groups, while galaxies in
  H~{\sc i}-poor groups clearly populate the lower end of the SSFR
  values. The intermediate groups span the range of SSFRs and populate
  the lower and somewhat higher values of SSFRs. Galaxies residing in
  H~{\sc i}-rich or intermediate groups tend to be MIR active, UV
  blue, and have higher SSFRs. Likewise, galaxies in H~{\sc i}-poor
  groups tend to be MIR inactive, UV red, and have lower SSFRs.}

\item{The UV-optical colours in Figure \ref{hammer_ssfr} show that
  MIR- quiescent galaxies lie along the red sequence, and we further
  see that these galaxies all have low SSFRs. The blue cloud is
  populated with MIR-active galaxies and these all have higher SSFRs
  (similar to \citet{walker13}). The green valley region
  defined by \citet{wyder07} is well populated with galaxies with
  intermediate SSFRs. The MIR canyon galaxy HCG 37E, is found to be in
  the green valley.}

\end{itemize}

In conclusion, the MIR colours, the UV colours and SSFRs all tell a
consistent story about galaxy evolution in the compact group
environment.  It is the SSFRs, not the SFRs, that drive the UV and MIR
colours. We see that in general MIR-active galaxies are also UV-blue,
have higher SSFRs and tend to lie in H~{\sc i}-rich or intermediate
groups. MIR-quiescent galaxies tend to be UV-red and have lower SSFRs,
and lie in H~{\sc i}- intermediate or poor groups. The persistent
presence of a bimodality in the SSFR tracers supports the idea that
the CG environment accelerates the evolution of galaxies from a state
of active star formation, to quiescence.

The next steps required to further understand galaxy evolution in
compact groups will involve investigating the spatial distributions of
star formation, stellar mass and specific star formation rate
distributions. \\

This work was supported by the Natural Science and Engineering
Research Council and the Ontario Early Researcher Award Program (LL,
SG). We thank the anonymous referee for useful comments which improved
the presentation of this work. We also thank T. Bitsakis, S. Rahmani
and N. Vulic for helpful discussions. This research has made use of
the NASA/IPAC Extragalactic Database (NED) which is operated by the
Jet Propulsion Laboratory, California Institute of Technology, under
contract with the National Aeronautics and Space Administration. This
research has made use of the VizieR catalogue access tool, CDS,
Strasbourg, France. The original description of the VizieR service was
published in \citet{ochsenbein00}.

\bibliographystyle{mn2e}
\bibliography{Paper1_MNRAS}

\label{lastpage}

\end{document}